\documentclass[%
reprint,
amsmath,amssymb,
aps,
pra,
]{revtex4-2}

\usepackage{graphicx}
\usepackage{dcolumn}
\usepackage{bm}

\usepackage[english]{babel}
\makeatletter
\let\ORIbbl@fixname\bbl@fixname
\def\bbl@fixname#1{%
	\@ifundefined{languagealias@\expandafter\string#1}
	{\ORIbbl@fixname#1}
	{\edef\languagename{\@nameuse{languagealias@#1}}}%
}
\newcommand{\definelanguagealias}[2]{%
	\@namedef{languagealias@#1}{#2}%
}
\makeatother
\definelanguagealias{en}{english}
\definelanguagealias{EN}{english}
\definelanguagealias{fr}{english}

\begin{document}


\title{High-resolution spectroscopy of arbitrary light sources using frequency combs}

\author{David Burghoff}
\affiliation{%
 Department of Electrical Engineering, University of Notre Dame, Notre Dame, IN 46656\\
}%

%
%

\date{\today}

\begin{abstract}
Multiheterodyne techniques using frequency combs---light sources whose lines are perfectly evenly-spaced---have revolutionized optical science. By beating an unknown signal with the many lines of a comb, its spectrum is recovered. However, these techniques have been restricted to measuring coherent sources, such as lasers. In this work, we demonstrate a new multiheterodyne technique that allows for nearly \textit{any} complex broadband spectrum to be retrieved using a comb. Two versions are introduced: a delayed comb technique that uses a tunable delay element, and a dual comb technique that uses a second comb. In each case, the spectrum of the source is recovered by Fourier transforming the correlation between two spectrograms. This approach is statistical in nature and is general to nearly any source (coherent or incoherent), allowing for the entire spectrum to be rapidly measured with high resolution.

\end{abstract}

\pacs{Valid PACS appear here}
\maketitle


\section{\label{sec:level1}Introduction}

A fundamental problem in optical science is detecting the spectrum of a remote source. Heterodyne detection is a powerful technique for performing high-resolution spectroscopy. By beating an unknown optical signal from a remote source with a known local oscillator (LO), one can measure the mixing between the two to achieve a high-precision measurement of the unknown signal’s spectrum. As this technique is particularly relevant for measuring the spectra of distant objects, it has found significant use in astronomy. However, a limitation of this technique is that it can only measure spectra over a limited bandwidth, effectively covering only frequencies that are within the mixer’s intermediate frequency (IF) bandwidth. This restriction can be overcome by requiring widely tunable local oscillators, but tuning can be difficult when the LO is an optical source, as tunable LOs typically require moving parts.

A potentially attractive alternative is the use of frequency combs, broadband light sources whose lines are perfectly evenly spaced. Frequency combs have enabled a wide variety of \textit{multiheterodyne techniques}, which in effect use multiple LO lines to detect multiple signals \cite{picqueFrequencyCombSpectroscopy2019}. For example, dual comb spectroscopy \cite{keilmannTimedomainMidinfraredFrequencycomb2004,coddingtonDualcombSpectroscopy2016,wangHighresolutionMultiheterodyneSpectroscopy2014,villaresDualcombSpectroscopyBased2014,yangTerahertzMultiheterodyneSpectroscopy2016,yuSiliconchipbasedMidinfraredDualcomb2018,suhMicroresonatorSolitonDualcomb2016} can be used to measure the spectrum of another comb, comb-referenced approaches can measure the spectrum of a laser \cite{delhayeFrequencyCombAssisted2009,bartaliniFrequencyCombAssistedTerahertzQuantum2014}, and vernier spectroscopies \cite{giorgettaFastHighresolutionSpectroscopy2010,coddingtonCharacterizingFastArbitrary2012,yangVernierSpectrometerUsing2019} can be used to measure the spectra of multi-line lasers. However, all of these techniques require that the resulting spectra do not overlap at intermediate frequencies (IFs), as overlapping IFs create an unavoidable ambiguity in the spectrum. This precludes their use in applications where the optical signal is broadband and may even be broader than the comb spacing---remote sensing, astronomy, biological systems, etc.

We introduce a new high-resolution multiheterodyne technique that is able to unravel the spectrum of \textit{arbitrary} light sources, even incoherent sources whose linewidths are much greater than the comb spacing. Inspired by the interferometer-based techniques that disambiguate individual comb lines \cite{mandonFourierTransformSpectroscopy2009,hugiMidinfraredFrequencyComb2012,burghoffTerahertzLaserFrequency2014,burghoffEvaluatingCoherenceTimedomain2015,hanSensitivitySWIFTSpectroscopy2020}, we show that a signal mixing with a comb can be fully disambiguated using a variable delay element, even when multiple signals appear at the same IF. We also show that a dual comb version of this approach can accomplish the same task, rapidly measuring the spectrum of any source without moving parts. Each version of this measurement has analogues to Fourier spectroscopy and preserves many of its features, such as the throughput and multiplex advantages \cite{fellgettTheoryInfraredSensitivities1951,jacquinotCaracteresCommunsAux1958}. Though the approach relies on the comb structure, it does \textit{not} require that the combs have a particular phase profile---the combs can be pulsed \cite{giorgettaFastHighresolutionSpectroscopy2010,yangVernierSpectrometerUsing2019,hillbrandInPhaseAntiPhaseSynchronization2020,bagheriPassivelyModelockedInterband2018} or not \cite{burghoffEvaluatingCoherenceTimedomain2015,singletonEvidenceLinearChirp2018,dongPhysicsFrequencymodulatedComb2018,henryPseudorandomDynamicsFrequency2017,cappelliRetrievalPhaseRelation2019,hillbrandInPhaseAntiPhaseSynchronization2020,delhayePhaseStepsResonator2015,metcalfHighPowerBroadlyTunable2013}.

\section{Overview}

\begin{figure}
	\includegraphics[width=\linewidth]{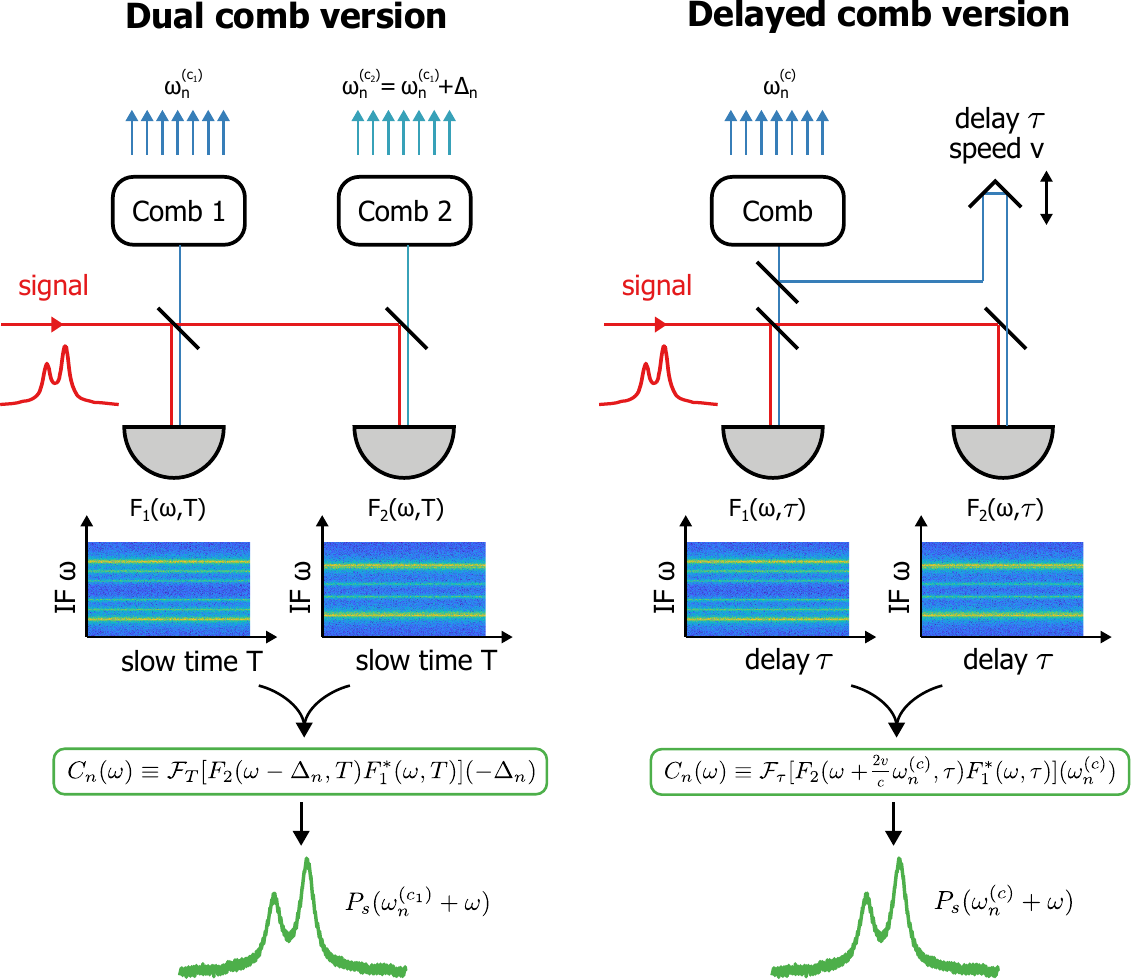}
	\caption{\label{fig:overview}Overview of the two approaches. In the dual comb version two combs are independently beat with the signal; in the delayed comb version one comb is split and delayed. Both detectors' spectrograms are computed and are correlated to reproduce the original signal's spectrum to high resolution (the measurement time divided by the number of comb lines).}
\end{figure}

A simplified version of the requisite experimental setups are shown in Figure \ref{fig:overview}.
In each measurement, the source to be measured is split and is mixed with two combs. In the dual comb version, these two combs come from separate sources. They may or may not be mutually coherent. In the delayed comb version, the second comb is generated from the first by using a variable delay element. Essentially, it is a Doppler-shifted comb \cite{mandonFourierTransformSpectroscopy2009}. In each case, the detector signals are digitized and processed into complex spectrograms. (This is done by dividing the data into batches and computing a short-time Fourier transform.) The product of the two spectrograms is computed, and the result is then Fourier transformed again to achieve the final result. Even for a fully incoherent source this correlation function is proportional to the power of the signal (offset from $n$th comb line), which allows it to reconstruct essentially any source. The details of the two versions differ only slightly.

In the dual comb version, the complex spectrograms $F_i(\omega,T)$ are functions of the IF frequency $\omega$ and the time of each spectrogram $T$. We denote the respective position of the $n$th comb lines as $\omega_n^{(c_1)}$ and $\omega_n^{(c_2)}\equiv\omega_n^{(c_1)}+\Delta_n$, where $\Delta_n$ is the separation between corresponding lines. We denote the respective complex amplitudes as $E_n^{(c_1)}$ and $E_n^{(c_2)}$. We then correlate the two spectrograms and compute $C_n(\omega)$, the Fourier transform of the spectrogram correlation along $T$,
\begin{align*}
	C_n(\omega)\equiv \mathcal{F}_T[F_2(\omega-\Delta_n,T)F_1^*(\omega,T)](-\Delta_n).
\end{align*}
As we show in Appendix \ref{sec:derivation}, this function is statistically-related to the spectrum of the source $P_s$ by
\begin{align}\label{du_result}
	\left\langle C_n(\omega)\right\rangle=E_n^{(c_2)*}E_n^{(c_1)}P_s(\omega_n^{(c_1)}+\omega).
\end{align}
In other words, the spectrum near a comb line can be determined simply by dividing out the amplitude of the dual-comb beat signal $E_n^{(c_2)*}E_n^{(c_1)}$. This result holds for both positive and negative IF frequencies as well as overlapping IF frequencies, allowing for complete disambiguation of the signal.

The delayed comb version is similar. Here the complex spectrograms $F_i(\omega,\tau)$ are functions of delay $\tau$, and are typically related to laboratory time by $\tau=\frac{2v}{c} T$ (provided the delay element moves at a constant velocity $v$). In this case there is only one set of comb frequencies and amplitudes, and the correct definition for $C_n$ is similar:
\begin{align}
	C_n(\omega)&\equiv \mathcal{F}_\tau[F_2(\omega+\tfrac{2v}{c}\omega_n^{(c)},\tau)F_1^*(\omega,\tau)](\omega_n^{(c)})\nonumber \\
	\left\langle C_n(\omega)\right\rangle&=E_n^{(c)*}E_n^{(c)}P_s(\omega_n^{(c)}+\omega). \label{de_result}
\end{align}
Once again, the spectrum near a comb line can be determined by dividing out the corresponding comb tooth power, in this case $|E_n^{(c)}|^2$. Detailed derivations are given in Appendix \ref{sec:derivation}. 

Both of these approaches are extremely general, reconstructing the source in practically all cases. The sole situation in which they will not correctly reproduce the spectrum of the signal is when there exist frequencies for which $\left\langle E_s(\omega)E_s^*(\omega+n\omega_r)\right\rangle\neq0$ over the duration of the measurement (where $\omega_r$ is the repetition rate of a comb and n is an integer). Over sufficiently long timescales, this will only fail when the source under consideration is deliberately chosen to match the combs' repetition rates, for example by attempting to measure another comb with the same spacing. The approach is also general for all types of combs, irrespective of the phase of the comb lines.

\section{Results}

As a relevant example, we consider a complex terahertz spectrum consisting of several lines, similar to the type of signal that is highly relevant for astronomy (for example, in measuring the spectral line energy distribution of carbon monoxide \cite{mashianHighJCOSLEDs2015}). We consider signals in the range of 4 to 5 THz and consider the dual comb version of the measurement. Our combs are assumed to span 4-5 THz with repetition rates of 10 GHz (typical parameters for quantum cascade laser combs \cite{burghoffTerahertzLaserFrequency2014,faistQuantumCascadeLaser2016}). Our signals are fairly broadband---with 100 MHz full-width half maximums (FWHMs)---and are generated numerically using a phase random walk process. A list of the line strengths and locations is shown in Table \ref{tab:table1}.

\begin{figure}
	\includegraphics[width=\linewidth]{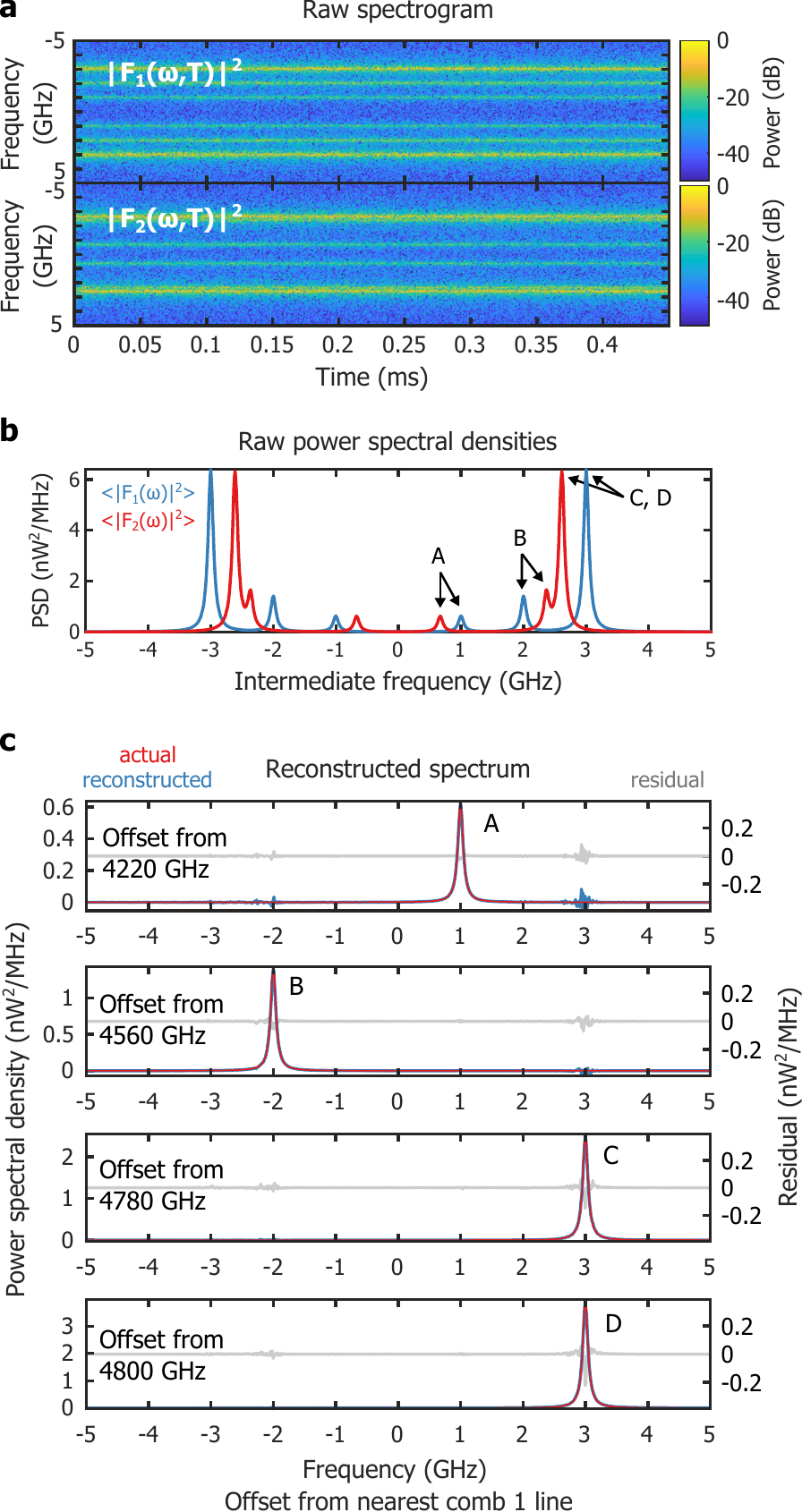}
	\caption{\label{fig:analysis}a. Magnitudes of the recorded spectrograms as a function of slow time and IF frequency (10 MHz RBW, 0.45 ms measurement time). b. Raw signal power spectral densities, with contributions from beating with various lines labeled. c. Reconstructed signals calculated from equation (\ref{du_result}), along with the actual spectrum.}
\end{figure}

\begin{table}[b]
	\begin{ruledtabular}
		\begin{tabular}{lp{15mm}p{15mm}p{15mm}p{15mm}}
			\textrm{Line}&
			\textrm{Frequency (GHz)}&
			\textrm{Power (pW)}&
			\textrm{Offset from comb 1 (GHz)}&
			\textrm{Offset from comb 2 (GHz)}\\
			\colrule\\[-0.2cm]
			A & 4211 & 0.1 & 	   1&0.669\\
			B & 4558 & 0.225 & -2&-2.366\\
			C & 4783 & 0.4 & 3&2.612\\
			D & 4803 & 0.625 & 3&2.610\\
		\end{tabular}
	\end{ruledtabular}	\caption{\label{tab:table1}%
	Lines of the spectrum considered. Comb 1 spans 4-5 THz with a repetition rate of 10 GHz. Comb 2 has a repetition rate of 10 GHz+1 MHz and has an additional offset of 0.3 GHz. Comb lines have a power of 1 mW per tooth.
}
\end{table}
These lines are chosen to illustrate the power of the technique. Lines A, C, and D appear at positive IFs (relative to the nearest comb line), while line B appears at a negative IF. With a single LO line and one detector, it is impossible to distinguish positive IFs from negative IFs. Furthermore, lines C and D appear at the exact same IF, which means that distinguishing them is impossible with all prior vernier-like techniques. In this case, the lines are relatively broadband but are much narrower than the comb spacing. The corresponding magnitude of the two spectrograms is shown in Figure \ref{fig:analysis}a.
Individually, the two spectrograms appear as a noisy version of the average power spectral density of the signal on the two detectors. This 'noise' actually arises from the incoherent nature of our signals, and it is the hidden correlations of these two signals that give rise to our computed result. Similarly, the average power spectral densities of the two signals are shown in Figure \ref{fig:analysis}b. They contain peaks from all lines beating with the combs---the result is a complex zoo of overlapping spectra.

Figure \ref{fig:analysis}c shows the results of our correlation calculation. For each comb line, we compute the real part of $C_n(\omega)$ and plot the result for comb lines that are near signal lines. We also show a theoretical prediction of $\left|E_n^{(c_2)*}E_n^{(c_1)}\right|P_s(\omega_n^{(c_1)}+\omega)$. The agreement between the two is excellent. For example, looking at the spectrum near line A (the weakest line), very little evidence of other lines is present despite the fact that much larger signals are beating at $\pm$2 GHz and $\pm$3 GHz on the raw spectrograms. Only small zero-mean image noise remains in the result. Note also that line B correctly appeared at -2 GHz, not 2 GHz. Though $F_1$ and $F_2$ must be conjugate-symmetric, $C_n$ is not. Finally, note that lines C and D are also correctly distinguished despite the fact that they overlap entirely with respect to both combs 1 and 2.

\begin{figure}
	\includegraphics[width=\linewidth]{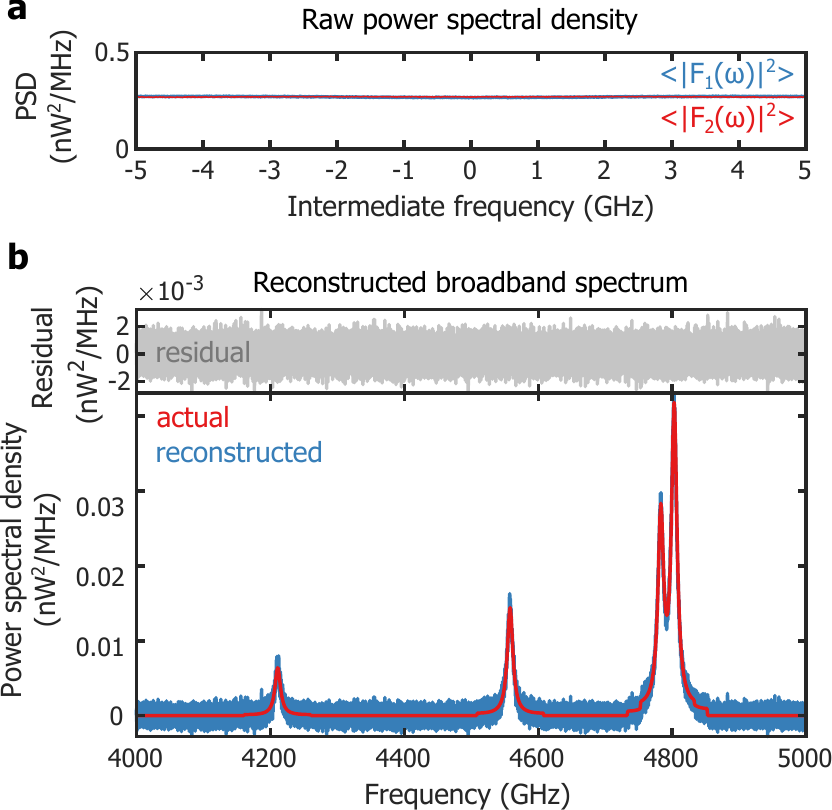}
	\caption{\label{fig:broadband}Reconstruction of a spectrum whose components are broader than the comb spacing. a. Raw power spectral densities of the recorded signals. They are essentially flat, having no discernible peaks. b. Spectral reconstruction and true spectrum (10 MHz RBW, 9 ms measurement time).}
\end{figure}

These results are valid for any source, even extremely incoherent sources. To illustrate this, we increase the FWHM of the lines shown above to 10 GHz---the comb repetition rate---and plot the reconstructions. With linewidths this broad, the entire IF span is filled with a relatively flat spectrum (see Figure \ref{fig:broadband}a). Since the true spectrum does not have narrow features, we plot the full reconstructed spectrum in Figure \ref{fig:broadband}b. Once again, the results are in good agreement with the theoretical prediction. Note that because lines C and D are only $2f_r$ apart, their lineshapes overlap both in real frequency and in intermediate frequency. Although in this case there is very little information to be gained from going to resolution bandwidths this narrow, in the general case where the spectrum's features are totally unknown the ability to perform high-resolution measurements is critical.
\section{Discussion}

As this technique extends multiheterodyne spectroscopy to the case of incoherent spectra, it maintains many of the appealing features of other multiheterodyne techniques. The dual comb version is similar to other dual comb techniques in that the ultimate time resolution of this measurement is determined by the ability to resolve different dual comb beat signals \cite{giorgettaFastHighresolutionSpectroscopy2010,coddingtonCharacterizingFastArbitrary2012}. As such, its time resolution is $1/\Delta f_r$ and can easily be within the range of microseconds (contingent on the signal-to-noise ratio). While the delayed comb version is limited by the mechanical speed of the delay element, chip-scale combs such as quantum cascade lasers and microresonator combs typically have repetition rates in the ten of GHz range, meaning that resolving their features requires only a travel distance of a few millimeters. This is conceivably within the range of 100 ms.

Next, we consider resolution. For perfect comb sources the resolution bandwidth (RBW) is determined by the batch length according to RBW$=\frac{1}{N_t \Delta t}$, where $\Delta t$ is the sample rate and $N_t$ is the number of samples per batch. However, resolving the dual comb beat term requires that there be at least as many batches ($N_T$) as the number of comb lines ($N_c$). If $T$ is the total measurement time $T\equiv N_T N_t \Delta t$, then this requires that $\textrm{RBW}>\frac{N_c}{T}$. In other words, the resolution is limited by the measurement time divided by the number of comb lines. This is worse than what is achievable with isolated lines (which allows for resolutions of $1/T$ \cite{giorgettaFastHighresolutionSpectroscopy2010}) because knowing that the line is isolated requires prior knowledge of the signal. Effectively, we traded some resolution for the ability to distinguish lines that overlap in the IF. Practically speaking this is only relevant for phase-stable combs; for free-running combs it is typically the linewidth of the combs themselves that set the technique's resolution.

In terms of sensitivity, this approach has many similarities to traditional Fourier spectroscopy. Because the entire signal is measured at once and is not demultiplexed, the multiplex advantage \cite{fellgettTheoryInfraredSensitivities1951} is maintained. Additionally, because the system does not require a single-mode source, the throughput advantage \cite{jacquinotCaracteresCommunsAux1958} is maintained. A detailed discussion is given in Appendix \ref{sec:sensitivity}; for additive noise, the total noise is the sum of image noises from every signal sharing an IF frequency. When compared with a tunable LO of the same total power, in best case scenario of non-overlapping signals the sensitivities are identical. In the worst case scenario of broadband spectra the RMS sensitivity is $\sqrt{N_c}$ worse, as the LO power of the individual lines is effectively divided by $N_c$. While this is problematic for certain astronomical applications where quantum-limited sensitivity is required, for dynamically-varying sources the dual comb version of this approach will be considerably faster than a widely-tunable LO (which typically requires moving parts) or a Fourier spectrometer. Detector nonlinearity will further degrade the sensitivity, see Appendix \ref{sec:nonlinearity}.

Like all comb-based spectroscopies, this approach suffers in the presence of comb phase noise. While the delayed comb version is fairly immune to such effects, the dual comb version is highly susceptible. Still, like dual-comb spectroscopy \cite{ideguchiAdaptiveRealtimeDualcomb2014,burghoffComputationalMultiheterodyneSpectroscopy2016,hebertSelfcorrectedChipbasedDualcomb2017,sterczewskiComputationalCoherentAveraging2019,sterczewskiComputationalDopplerlimitedDualcomb2019,burghoffGeneralizedMethodComputational2019}, it can be corrected after the fact. This procedure is detailed in Appendix \ref{sec:phase_correction}, and essentially requires that a standard dual comb measurement (measuring the beating of the two combs) is performed. This additional measurement can also be used to calibrate the amplitude, as a dual comb measurement will produce the beat signal $E_n^{(c_2))}E_n^{(c_1))*}$ that is needed to normalize the result.

Lastly, we place this result in the context of earlier work, in particular with respect to vernier spectroscopy \cite{giorgettaFastHighresolutionSpectroscopy2010,coddingtonCharacterizingFastArbitrary2012,yangVernierSpectrometerUsing2019}. While the delayed comb version of this approach has not been demonstrated in any capacity (to our knowledge), the dual comb version is experimentally very similar to vernier spectroscopy. Previous work did not attempt to perform spectroscopy of incoherent sources, focusing on swept-source diode lasers \cite{giorgettaFastHighresolutionSpectroscopy2010,coddingtonCharacterizingFastArbitrary2012,yangVernierSpectrometerUsing2019} and fiber mode-locked lasers \cite{yangVernierSpectrometerUsing2019}. The key distinction is that in all of these cases, individual lines can be isolated provided when they do not occur at the same IF frequency, and correlation maximization procedures can be used to find the comb order number of each line. However, this approach does not generalize to arbitrary incoherent spectra, as continuous overlapping spectra from different orders are not compatible with this approach. Still, equation (\ref{du_result}) is immediately applicable to these results, providing a new way to analyze them.

\section{Conclusion}

We have shown theoretically and numerically that frequency combs can be used to unravel the spectrum of an arbitrary incoherent light source. This can be done either use two separate combs or one comb with a copy that has been delayed. Even when the beating of the two combs appears random and chaotic, a periodically-varying correlation persists in the complex spectrograms, and this can be exploited to infer the spectrum. This result is compatible with all existing comb technologies and will allow for passive high-speed spectroscopy of essentially any dynamically-varying electromagnetic source. For example, one can imagine applications in reaction kinetics \cite{pinkowskiDualcombSpectroscopyHightemperature2020}, in biology \cite{karpovPhotonicChipbasedSoliton2018,klockeSingleShotSubmicrosecondMidinfrared2018}, and even in millimeter-wave systems \cite{hoffmannRealTimeLowNoiseUltrabroadband2011}. 

\appendix

\section{Detailed derivation}\label{sec:derivation}

To show that equations (\ref{du_result}) and (\ref{de_result}) hold, we will derive the result for the dual comb version first and extend it to the delayed comb version. All electric fields are expressed as a superposition of exponentials as
\begin{align*}
	E_{c_i}(t)=\sum_n E_n^{(c_i)} e^{i\omega_n^{(c_i)}t}\textrm{ and }E_{s}(t)=\sum_m E_n^{(s)} e^{i\omega_m^{(s)}t}
\end{align*}
where $E_{c_i}(t)$ is the field of the ith comb and $E_{s}(t)$ is the field of the signal to be measured. For convenience the signal is represented as a summation rather than an integral. In a heterodyne measurement, the raw signal that is recorded is
\begin{align*}
	S_i(t)&=\frac{1}{2} \left|E_{c_i}(t)+E_s(t)\right|^2 \sim E_{c_i}^* E_s \\
	&= \sum_{n,m} E_n^{(c_i)*} E_m^{(s)} e^{i(\omega_m^{(s)}-\omega_n^{(c_i)})t}
\end{align*}
where we neglect the intracomb beat terms (which occur only at vanishingly-narrow multiples of the repetition rate) and the intrasignal beat terms (which are assumed to be small).

We assume that our detector has a bandwidth larger than $f_r/2$ (half the comb repetition rates) and that the signal is digitized with a sample period $\Delta t$ sufficiently small to avoid aliasing. In the spirit of spectrograms, time is divided into two separate time axes---the fast time $t_i$ and the slow time $T_j$---such that the total time is given by $t=t_i+T_j$. The data is similarly divided into batches that are $N_t \Delta t$ long (which determines the resolution bandwidth). In other words, $T_j$ can be taken as $T_j=(j-1)N_t \Delta t$, where j is an integer.

\vspace{0.2cm}\textit{Dual comb version. } First, we calculate the signals in terms of the FFT IF frequencies $\omega_k \equiv \frac{2\pi}{N_t\Delta t}k$. For the FFT, we use the convention that $\mathcal{F}[f](\omega_k)\equiv\frac{1}{N_t}\sum_i e^{-i\omega_k t_i}f_i$. For comb 1, we find that the spectrogram $F_1(\omega_k,T_j)$ is given by
\begin{align*}
	F_1(\omega_k,T_j)&=\frac{1}{N_t}\sum_i e^{-i\omega_k t_i}S_1(t_i+T_j) \\
	&=\frac{1}{N_t}\sum_{i,n,m}E_n^{(c_1)*}E_m^{(s)}e^{i(\omega_m^{(s)}-\omega_n^{(c_1)}-\omega_k)t_i}\\[-1em]&\hspace{2.9cm}\times e^{i(\omega_m^{(s)}-\omega_n^{(c_1)})T_j}.
\end{align*}
Because our resolution bandwidth is determined by our batch length (RBW$=\frac{1}{N_t \Delta t}$), in order to proceed we make an approximation in which our signal frequencies are all an integer number of RBWs away from comb 1, i.e. $\omega_m^{(s)}-\omega_n^{(c_1)} = \frac{2\pi}{N_t \Delta t} l$ for some integer $l$. As a result, $e^{i(\omega_m^{(s)}-\omega_n^{(c_1)})T_j}=1$, and
\begin{align*}
	F_1(\omega_k,T_j)&=\sum_n E_n^{(c_1)*} \frac{1}{N_t} \sum_i e^{-i(\omega_n^{(c_1)}+\omega_k)t_i} \nonumber \\[-0.2em] &\hspace{2.5cm}\times \sum_m E_m^{(s)}e^{i\omega_m^{(s)}t_i}\nonumber \\
	&= \sum_n E_n^{(c_1)*} E_s(\omega_n^{(c_1)}+\omega_k),
\end{align*}
where in the last line we used the definition of the FFT twice. Due to our finite resolution bandwidth approximation, the spectrogram would appear to be constant in $T$. However, once this approximation has been made it \textit{cannot }be modified when computing the same quantity for comb 2. Performing a $1\rightarrow2$ substitution, one finds that
\begin{align*}
	e^{i(\omega_m^{(s)}-\omega_n^{(c_2)})T_j}&=e^{i(\omega_m^{(s)}-\omega_n^{(c_1)}-\Delta_n)T_j}=e^{-i\Delta_n T_j}\nonumber\\
	F_2(\omega_k,T_j)&= \sum_n E_n^{(c_2)*} E_s(\omega_n^{(c_2)}+\omega_k)e^{-i\Delta_n T_j}.
\end{align*}
Thus, while $F_1$ is stationary in time, $F_2$ is not, and in fact beats periodically at the dual comb frequencies $\Delta_n$. Provided the data has been recorded long enough to resolve individual beat frequencies (i.e., that the data is at least recorded for $1/\Delta f_r=1/|f_{r_2}-f_{r_1}|$), these beatings can be resolved. For simplicity, we assume that the number of batches $N_T$ has been chosen to ensure that $N_T N_t \Delta t=1/\Delta f_r$, which will resolve exactly one dual comb beat tooth.

Finally, we compute the Fourier transform of the correlation function and its expectation value:
\begin{align*}
	C_n(\omega_k)&\equiv \mathcal{F}_T[F_2(\omega_k-\Delta_n,T_j)F_1^*(\omega_k,T_j)](-\Delta_n) \\
	\left\langle C_n(\omega_k)\right\rangle  &= \frac{1}{N_T} \sum_{j,l,m}E_l^{(c_2)*}E_m^{(c_1)} e^{i(\Delta_n-\Delta_l)T_j} \\[-0.3em]
	&\hspace{-0.5cm}\times \left\langle E_s(\omega_l^{(c_1)}+\Delta_l-\Delta_n+\omega_k)E_s^*(\omega_m^{(c_1)}+\omega_k)\right\rangle 
\end{align*}
Because the number of batches was chosen to be an integer number of $1/\Delta f_r$, the summation $\frac{1}{N_T}\sum_j e^{i(\Delta_n-\Delta_l)T_j}$ is a summation over roots of unity and vanishes unless $n=l$, leaving
\begin{align}
	\left\langle C_n(\omega_k)\right\rangle  &= \sum_{m}E_n^{(c_2)*}E_m^{(c_1)} \nonumber \\[-0.3em]
	&\hspace{0.5cm}\times \left\langle E_s(\omega_n^{(c_1)}+\omega_k)E_s^*(\omega_m^{(c_1)}+\omega_k)\right\rangle 
\end{align}
This result is general for any source. For the vast majority of sources we can make an additional assumption, which is that the long-term correlation between components of frequencies spaced by the repetition rate of the comb vanishes. This assumption is essentially valid for any source that is not a comb of the same spacing as either of the LO combs. With this additional assumption we can eliminate the cross terms, leaving
\begin{align}
	\left\langle C_n(\omega_k)\right\rangle  &= E_n^{(c_2)*}E_n^{(c_1)} \left\langle \left|E_s(\omega_n^{(c_1)}+\omega_k)\right|^2 \right\rangle. 
\end{align}
By calculating every $C_n$, the signal's power relative to a comb line can always be extracted.

\vspace{0.2cm}\textit{Delayed comb version.} For the delayed comb, the analysis is similar. The analysis for comb 1 is fully identical (letting $E_n^{(c_1)} \rightarrow E_n^{(c)}$), and the analysis for comb 2 is found by setting $E_n^{(c_2)} \rightarrow E_n^{(c)} e^{-i \omega_n^{(c)} \tau}$ and $\Delta_n=0$. However, a subtlety arises when delay is a linear function of lab time (i.e., a linear scan is performed rather than a step scan). Because delay changes during the batch, this has the same effect as Doppler shifting the IF frequencies in a manner similar to a nonzero $\Delta_n$. One should therefore proceed as before, but using the explicit mapping $\tau\rightarrow\frac{2v}{c}(t_i+T_j)$. This results in
\begin{align*}
	F_2(\omega_k,T_j) = \sum_n E_n^{(c)*} E_s\left(\omega_n^{(c)}-\omega_n^{(c)}\frac{2v}{c}+\omega_k\right) e^{i\omega_n^{(c)}\frac{2v}{c} T_j}
\end{align*}
By comparing this to the dual comb version of the same result, we find that $-\omega_n^{(c)}\frac{2v}{c}$ has replaced $\Delta_n$. If we now define $\tau_j\equiv \frac{2v}{c}T_j$ and make the appropriate substitutions into our definition of $C_n$, we find that we must instead calculate
\begin{align*}
	C_n(\omega_k)&\equiv \mathcal{F}_\tau[F_2(\omega_k+\omega_n^{(c)}\frac{2v}{c},\tau_j)F_1^*(\omega_k,\tau_j)](\omega_n^{(c)}) 
\end{align*}
which results in
\begin{align*}
	\left\langle C_n(\omega_k)\right\rangle &= \left|E_n^{(c)}\right|^2 \left\langle \left| E_s(\omega_n^{(c)}+\omega_k) \right|^2\right\rangle.
\end{align*}

\section{Sensitivity}\label{sec:sensitivity}
\begin{figure}
	\includegraphics[width=\linewidth]{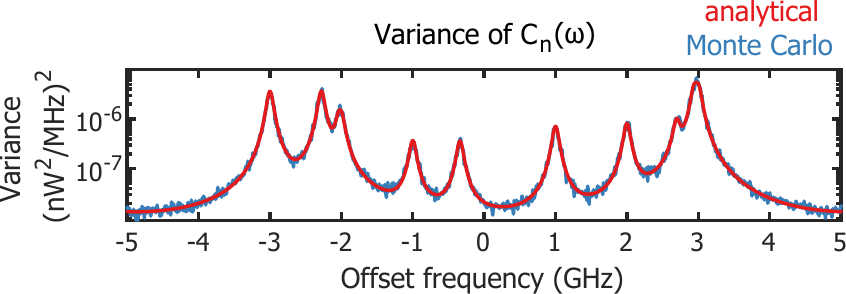}
	\caption{\label{fig:variance}Variance of the reconstruction in Fig. \ref{fig:analysis} for white noise of RMS power 10 nW, evaluated analytically using (\ref{var_eqn}) and numerically using Monte Carlo.}
\end{figure}

For astronomical applications, it is important to analyze the sensitivity of this approach. We do so for the case of additive white Gaussian noise (AWGN) and compared with the sensitivity of a tunable local oscillator. We assume that each detector measurement $S_i(t)$ is perturbed by a white noise source with variance $\sigma^2$. This noise source is also taken to be uncorrelated between each detector. By propagating this noise through the reconstruction, one can show that the variance of the extracted signal in the dual comb version is given by
\begin{align}
\textrm{Var}\left[\textrm{Re}\,C_n(\omega)\right]=\frac{1}{2N_T}&\left(\frac{\sigma^2}{N_t}\right)^2 \nonumber\\
+\frac{\sigma^2}{2N_T N_t} \sum_m &\left|E_m^{(c_1)}\right|^2 P_s(\omega_m^{(c_1)}+\omega) \nonumber\\[-0.3cm] +&\left|E_m^{(c_2)}\right|^2 P_s(\omega_m^{(c_2)}-\Delta_n+\omega)\label{var_eqn}
	\end{align}
As an example, in Fig. \ref{fig:variance} we plot this expression for the data plotted in Fig. \ref{fig:analysis}.
Because the summation over m is over both positive and negative frequencies, this means that the noise at a single frequency comes from the double-sided power spectral density of \textit{all} signal components that share an IF with the frequency under consideration. The noise is frequency-dependent and resembles the sum of the two raw power spectral densities. This is practically speaking a dynamic range limitation: if one attempts to measure a weak signal that shares an IF with a much stronger signal, image noise generated by the stronger signal will swamp the weaker one. Whether this is tolerable depends on the source and on the application. For spectra consisting of relatively narrow lines it can be avoided by a proper choice of $f_r$, for example. 

The corresponding expression for using a single LO with electric field $E_0 e^{i\omega_0 t}+c.c.$ to measure double sideband intensity is
\begin{align*}
	\textrm{Var} \left[I_0(\omega_0)\right] &= \frac{1}{N_T}\left(\frac{\sigma^2}{N_t}\right)^2 \nonumber\\
	&+\frac{2\sigma^2}{N_T N_t} \left|E_0\right|^2 \left(P_s(\omega_0+\omega)+P_s(\omega_0-\omega)\right)
\end{align*}
Both expressions have a constant term independent of the signal---usually neglected since it is fourth-order in the noise---and both have a term that decreases linearly with the total measurement time. To compare these two expressions, we note that for a uniform comb the measurement time of the tunable LO must be $N_c$ times smaller to account for the fact that the comb measurement is multiplexed. In addition, the power per comb tooth for the comb version should be $N_c$ times smaller than the tunable LO's power if the mixer is to be optimally pumped. When there are M overlapping lines in the IF, the signal-to-noise ratio (SNR) for both the comb and the tunable LO are respectively given by
\begin{align*}
	\textrm{SNR}_\textrm{comb}^2 &= \frac{1}{M}\frac{N_T N_t}{N_c} \frac{1}{2\sigma^2} \left|E_0\right|^2 P_s(\omega_n^{(c_1)}+\omega)\\
	\textrm{SNR}_\textrm{tuned}^2 &= \hspace{1.3em}\frac{N_T N_t}{N_c} \frac{1}{2\sigma^2} \left|E_0\right|^2 P_s(\omega_0+\omega)
\end{align*}
In the best case scenario, where no lines overlap, the SNRs are identical. In the worst case scenario (a broadband light source), $M=N_c$ and the SNR is a factor of $\sqrt{N_c}$ times worse.

In addition to additive noise, there is multiplicative noise that arises from the fact that even incoherent signals can have transient frequency domain correlations. Even when $\left\langle E_s(\omega)E_s^*(\omega+n\omega_r)\right\rangle=0$, it will not be the case that $\left\langle \left| E_s(\omega)E_s^*(\omega+n\omega_r)\right|^2\right\rangle=0$. This manifests as noise, and it effectively limits the dynamic range of the measurement. While the exact form depends on the details of the source---white frequency noise differs from flicker noise, for example---the variance is typically proportional to equation (\ref{var_eqn}) for sufficiently broadband spectra.

\section{Phase correction}\label{sec:phase_correction}

\begin{figure}
	\includegraphics[width=\linewidth]{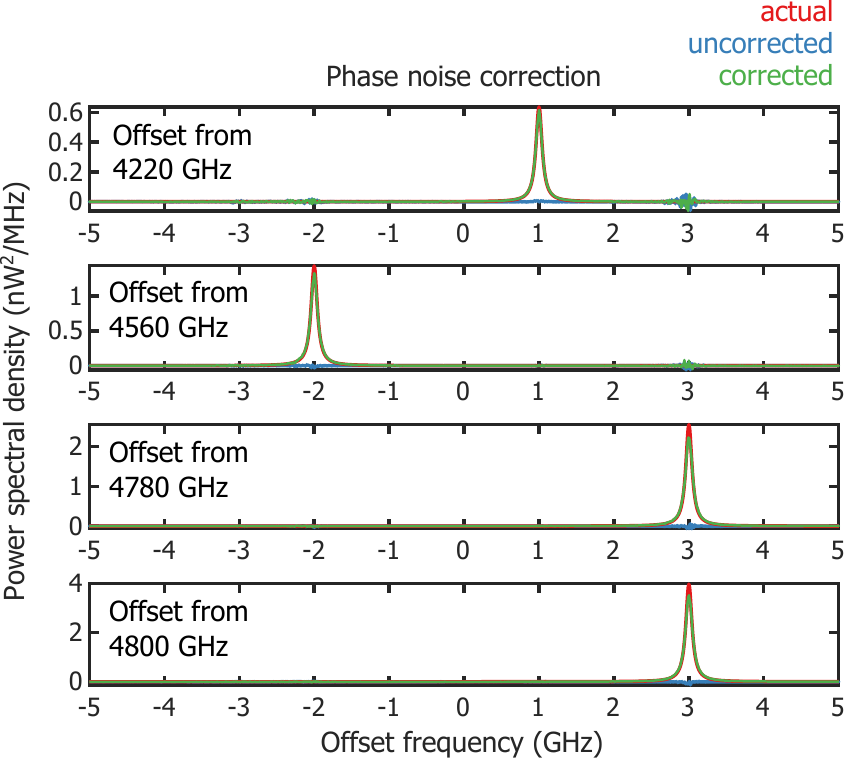}
	\caption{\label{fig:phase_noise} Spectral reconstruction in the presence of phase noise. Offset fluctuations of 1 MHz and repetition rate fluctuations of 1 kHz preclude reconstruction, but using the alternative correlation function correctly recovers the original spectra.}
\end{figure}

In the derivation of this approach we assumed that the comb lines being used to probe the signal were free of phase noise. In fact, this assumption can be relaxed if an additional dual comb spectroscopy measurement is performed to measure their mutual phase fluctuations and compensate for them. Suppose that the dual comb beating of pair n has been digitized and is given by $V_n(t)=E_n^{(c_2)}E_n^{(c_1)*}e^{i\Delta_n t}$, and that its magnitude has been divided out to construct $p_n(t)=e^{i(\phi_{2n}-\phi_{1n})}e^{i\Delta_n t}$. The correlation function is then given by
\begin{align*}
		C_n(\omega)&= \mathcal{F}_T[F_2(\omega-\Delta_n,T)F_1^*(\omega,T)](-\Delta_n)\\
		&= \frac{1}{N_T N_t}\sum_{i,j} e^{i\Delta_n (T_j+t_i)} e^{-i\omega t_i}S_2(t_i,T_j)F_1^*(\omega,T_j).
\end{align*}
Since the total time is given by $t=t_i+T_j$, the explicit dependence on $\Delta_n$ can be removed by substituting in $p_n e^{i(\phi_{1n}-\phi_{2n})}$ and noting that the summation over i is merely an FFT over the $t_i$ axis:
\begin{align*}
	C_n(\omega)&= \frac{1}{N_T N_t}\sum_{i,j} p_n e^{i(\phi_{1n}-\phi_{2n})} e^{-i\omega t_i}S_2(t_i,T_j)F_1^*(\omega,T_j) \\
	&= e^{i(\phi_{1n}-\phi_{2n})} \left\langle \mathcal{F}_t[p_n S_2](\omega,T)F_1^*(\omega,T)\right\rangle_T \\[-0.3cm]
\end{align*}
This expression is convenient since it eliminates any explicit references to $\Delta_n$ by premultiplying the signal before the spectrogram calculation. Not only does it remove phase noise, but it also makes calculation of the power spectrum more convenient, as it removes the global phase of the dual comb lines. By defining the alternative correlation function $\hat{C}_n(\omega)$ as
 \begin{align}
 	\hat{C}_n(\omega)&\equiv \left\langle \mathcal{F}_t[p_n S_2](\omega,T)F_1^*(\omega,T)\right\rangle_T \label{phase_corrected_correlation}
 \end{align}
we find that
\begin{align*}
	\left\langle \hat{C}_n(\omega)\right\rangle = \left|E_n^{(c_1)} E_n^{(c_2)} \right| P_s(\omega_n^{(c_1)}+\omega).
\end{align*}
This result guarantees that phase noise does not contribute to the reconstruction, essentially by eliminating the phase entirely. Figure \ref{fig:phase_noise} illustrates this for the data in Figure \ref{fig:analysis} for free-running combs.
The combs have random walk phase noise producing offset fluctuations with a 1 MHz FWHM and repetition rate fluctuations with a 1 kHz FWHM, similar to what is found in free-running QCLs. As a result, the reconstruction produced by the previous approach fails. However, using the alternative correlation defined in (\ref{phase_corrected_correlation}), the correct result is recovered.

\section{Detector nonlinearity}\label{sec:nonlinearity}

Because we are using many frequencies to reconstruct our signal, it is possible that detector/mixer nonlinearity could negatively impact the results of the measurement, for example by distorting the signal. For example, this can be a challenge in dual comb spectroscopy when highly nonlinear mixers such as hot electron bolometers are used to measure the spectrum \cite{yangTerahertzMultiheterodyneSpectroscopy2016}, as nonlinearity will cause different comb lines to mix. However, in this case we do not expect such effects to be significant. When operating in the heterodyne limit, the measured signal is given by
\begin{align*}
	S_i(t)&=\frac{1}{2} \left|E_{c_i}(t)+E_s(t)\right|^2 = \frac{1}{2} \left|E_{c_i}\right|^2 + E_{c_i}^* E_s + \frac{1}{2} \left|E_{s}\right|^2
\end{align*}
and the $\left|E_{c_i}\right|^2$ term is much larger than the heterodyne term. Even so, it is easy to ignore because it only beats at multiples of the repetition rate (i.e., is periodic). Any nonlinearity will act upon it only to produce another signal that beats at the repetition rate, and we can therefore expect to continue to be able to ignore it.

We therefore expect that the lone effect of nonlinearity on the heterodyne measurement is to reduce its sensitivity. For example, if a standard two-level saturation model is used to model the detector response, then the output signal will be related to the input power by
\begin{align*}
	S(t)=\frac{P(t)}{1+\frac{P(t)}{P_{sat}}}
\end{align*}
where $P_{sat}$ is the saturation power. The heterodyne responsivity is essentially the differential response of this model, which is squared since both detectors suffer this decrease:
\begin{align*}
	\left(\frac{dS}{dP}\right)^2=\left(1+\frac{P}{P_{sat}}\right)^{-4}
\end{align*}
Once again, we have verified this numerically.
\begin{figure}
	\includegraphics[width=\linewidth]{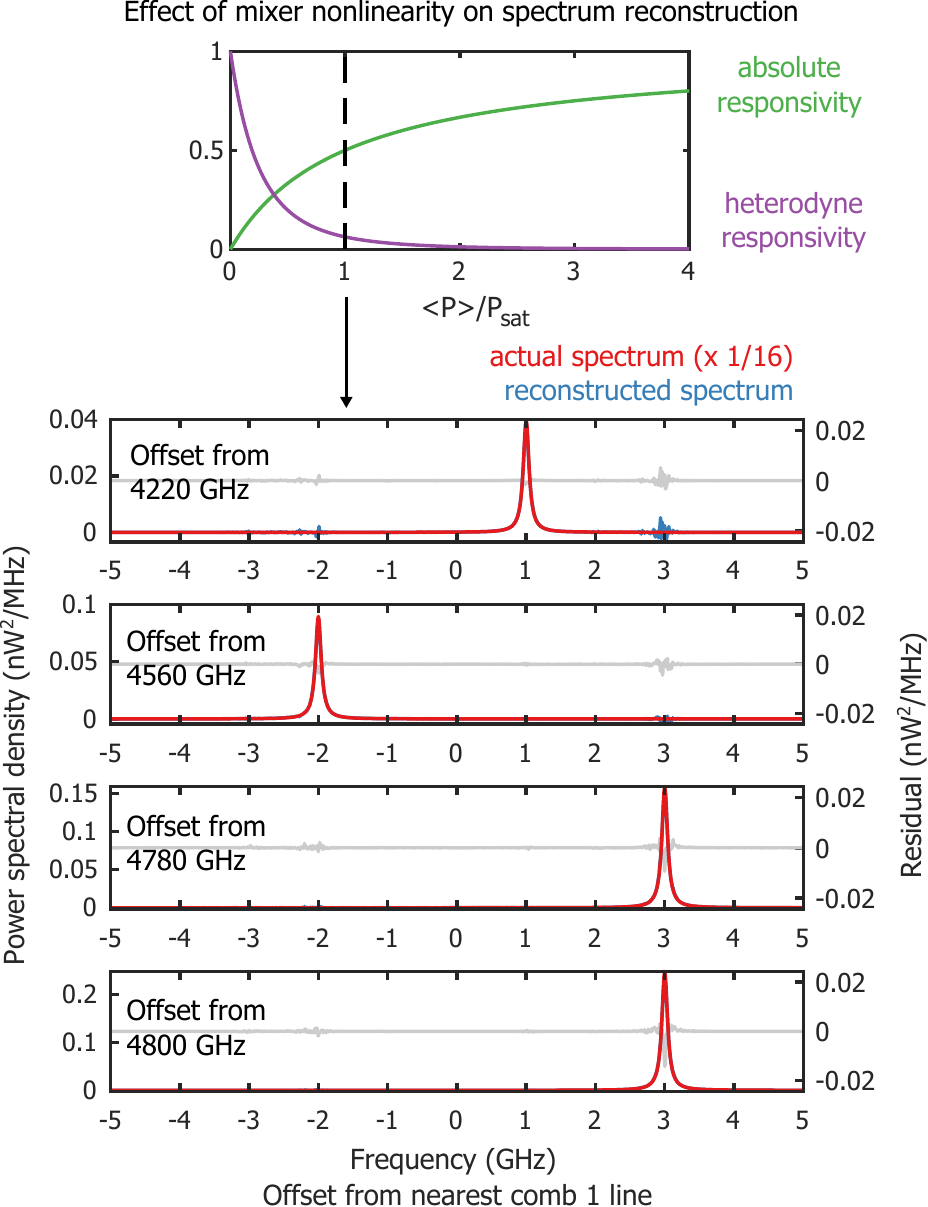}
	\caption{\label{fig:nonlinear} Spectral reconstruction in the presence of mixer nonlinearity. Although the reconstruction is a factor of 16 times lower due to a reduction in the heterodyne sensitivity, it is otherwise unaffected.}
\end{figure}
Figure \ref{fig:nonlinear} shows the result of passing the narrowband data from before through the two-level saturation model, choosing the saturation power to coincide with the average power $P_{sat}=\left\langle P(t)\right\rangle$. As expected, there is essentially no effect on the reconstruction, except that all measured reconstructed power spectral densities have been reduced by a factor of $2^4=16$. As with conventional heterodyne measurements, there will always be an optimal pumping power where the mixer is not yet saturated, but where the conversion gain is maximal.


\bibliography{zotero_library}

\begin{thebibliography}{40}%
\makeatletter
\providecommand \@ifxundefined [1]{%
 \@ifx{#1\undefined}
}%
\providecommand \@ifnum [1]{%
 \ifnum #1\expandafter \@firstoftwo
 \else \expandafter \@secondoftwo
 \fi
}%
\providecommand \@ifx [1]{%
 \ifx #1\expandafter \@firstoftwo
 \else \expandafter \@secondoftwo
 \fi
}%
\providecommand \natexlab [1]{#1}%
\providecommand \enquote  [1]{``#1''}%
\providecommand \bibnamefont  [1]{#1}%
\providecommand \bibfnamefont [1]{#1}%
\providecommand \citenamefont [1]{#1}%
\providecommand \href@noop [0]{\@secondoftwo}%
\providecommand \href [0]{\begingroup \@sanitize@url \@href}%
\providecommand \@href[1]{\@@startlink{#1}\@@href}%
\providecommand \@@href[1]{\endgroup#1\@@endlink}%
\providecommand \@sanitize@url [0]{\catcode `\\12\catcode `\$12\catcode
  `\&12\catcode `\#12\catcode `\^12\catcode `\_12\catcode `\%12\relax}%
\providecommand \@@startlink[1]{}%
\providecommand \@@endlink[0]{}%
\providecommand \url  [0]{\begingroup\@sanitize@url \@url }%
\providecommand \@url [1]{\endgroup\@href {#1}{\urlprefix }}%
\providecommand \urlprefix  [0]{URL }%
\providecommand \Eprint [0]{\href }%
\providecommand \doibase [0]{https://doi.org/}%
\providecommand \selectlanguage [0]{\@gobble}%
\providecommand \bibinfo  [0]{\@secondoftwo}%
\providecommand \bibfield  [0]{\@secondoftwo}%
\providecommand \translation [1]{[#1]}%
\providecommand \BibitemOpen [0]{}%
\providecommand \bibitemStop [0]{}%
\providecommand \bibitemNoStop [0]{.\EOS\space}%
\providecommand \EOS [0]{\spacefactor3000\relax}%
\providecommand \BibitemShut  [1]{\csname bibitem#1\endcsname}%
\let\auto@bib@innerbib\@empty
\bibitem [{\citenamefont {Picqu{\'e}}\ and\ \citenamefont
  {H{\"a}nsch}(2019)}]{picqueFrequencyCombSpectroscopy2019}%
  \BibitemOpen
  \bibfield  {author} {\bibinfo {author} {\bibfnamefont {N.}~\bibnamefont
  {Picqu{\'e}}}\ and\ \bibinfo {author} {\bibfnamefont {T.~W.}\ \bibnamefont
  {H{\"a}nsch}},\ }\bibfield  {title} {{\selectlanguage {en}\bibinfo {title}
  {Frequency comb spectroscopy}},\ }\href
  {https://doi.org/10.1038/s41566-018-0347-5} {\bibfield  {journal} {\bibinfo
  {journal} {Nature Photonics}\ }\textbf {\bibinfo {volume} {13}},\ \bibinfo
  {pages} {146} (\bibinfo {year} {2019})}\BibitemShut {NoStop}%
\bibitem [{\citenamefont {Keilmann}\ \emph {et~al.}(2004)\citenamefont
  {Keilmann}, \citenamefont {Gohle},\ and\ \citenamefont
  {Holzwarth}}]{keilmannTimedomainMidinfraredFrequencycomb2004}%
  \BibitemOpen
  \bibfield  {author} {\bibinfo {author} {\bibfnamefont {F.}~\bibnamefont
  {Keilmann}}, \bibinfo {author} {\bibfnamefont {C.}~\bibnamefont {Gohle}},\
  and\ \bibinfo {author} {\bibfnamefont {R.}~\bibnamefont {Holzwarth}},\
  }\bibfield  {title} {{\selectlanguage {en}\bibinfo {title} {Time-domain
  mid-infrared frequency-comb spectrometer}},\ }\href
  {https://doi.org/10.1364/OL.29.001542} {\bibfield  {journal} {\bibinfo
  {journal} {Optics Letters}\ }\textbf {\bibinfo {volume} {29}},\ \bibinfo
  {pages} {1542} (\bibinfo {year} {2004})}\BibitemShut {NoStop}%
\bibitem [{\citenamefont {Coddington}\ \emph {et~al.}(2016)\citenamefont
  {Coddington}, \citenamefont {Newbury},\ and\ \citenamefont
  {Swann}}]{coddingtonDualcombSpectroscopy2016}%
  \BibitemOpen
  \bibfield  {author} {\bibinfo {author} {\bibfnamefont {I.}~\bibnamefont
  {Coddington}}, \bibinfo {author} {\bibfnamefont {N.}~\bibnamefont
  {Newbury}},\ and\ \bibinfo {author} {\bibfnamefont {W.}~\bibnamefont
  {Swann}},\ }\bibfield  {title} {{\selectlanguage {en}\bibinfo {title}
  {Dual-comb spectroscopy}},\ }\href {https://doi.org/10.1364/OPTICA.3.000414}
  {\bibfield  {journal} {\bibinfo  {journal} {Optica}\ }\textbf {\bibinfo
  {volume} {3}},\ \bibinfo {pages} {414} (\bibinfo {year} {2016})}\BibitemShut
  {NoStop}%
\bibitem [{\citenamefont {Wang}\ \emph {et~al.}(2014)\citenamefont {Wang},
  \citenamefont {Soskind}, \citenamefont {Wang},\ and\ \citenamefont
  {Wysocki}}]{wangHighresolutionMultiheterodyneSpectroscopy2014}%
  \BibitemOpen
  \bibfield  {author} {\bibinfo {author} {\bibfnamefont {Y.}~\bibnamefont
  {Wang}}, \bibinfo {author} {\bibfnamefont {M.~G.}\ \bibnamefont {Soskind}},
  \bibinfo {author} {\bibfnamefont {W.}~\bibnamefont {Wang}},\ and\ \bibinfo
  {author} {\bibfnamefont {G.}~\bibnamefont {Wysocki}},\ }\bibfield  {title}
  {{\selectlanguage {en}\bibinfo {title} {High-resolution multi-heterodyne
  spectroscopy based on {{Fabry}}-{{Perot}} quantum cascade lasers}},\ }\href
  {https://doi.org/10.1063/1.4862756} {\bibfield  {journal} {\bibinfo
  {journal} {Applied Physics Letters}\ }\textbf {\bibinfo {volume} {104}},\
  \bibinfo {pages} {031114} (\bibinfo {year} {2014})}\BibitemShut {NoStop}%
\bibitem [{\citenamefont {Villares}\ \emph {et~al.}(2014)\citenamefont
  {Villares}, \citenamefont {Hugi}, \citenamefont {Blaser},\ and\ \citenamefont
  {Faist}}]{villaresDualcombSpectroscopyBased2014}%
  \BibitemOpen
  \bibfield  {author} {\bibinfo {author} {\bibfnamefont {G.}~\bibnamefont
  {Villares}}, \bibinfo {author} {\bibfnamefont {A.}~\bibnamefont {Hugi}},
  \bibinfo {author} {\bibfnamefont {S.}~\bibnamefont {Blaser}},\ and\ \bibinfo
  {author} {\bibfnamefont {J.}~\bibnamefont {Faist}},\ }\bibfield  {title}
  {{\selectlanguage {en}\bibinfo {title} {Dual-comb spectroscopy based on
  quantum-cascade-laser frequency combs}},\ }\href
  {https://doi.org/10.1038/ncomms6192} {\bibfield  {journal} {\bibinfo
  {journal} {Nature Communications}\ }\textbf {\bibinfo {volume} {5}},\
  \bibinfo {pages} {5192} (\bibinfo {year} {2014})}\BibitemShut {NoStop}%
\bibitem [{\citenamefont {Yang}\ \emph {et~al.}(2016)\citenamefont {Yang},
  \citenamefont {Burghoff}, \citenamefont {Hayton}, \citenamefont {Gao},
  \citenamefont {Reno},\ and\ \citenamefont
  {Hu}}]{yangTerahertzMultiheterodyneSpectroscopy2016}%
  \BibitemOpen
  \bibfield  {author} {\bibinfo {author} {\bibfnamefont {Y.}~\bibnamefont
  {Yang}}, \bibinfo {author} {\bibfnamefont {D.}~\bibnamefont {Burghoff}},
  \bibinfo {author} {\bibfnamefont {D.~J.}\ \bibnamefont {Hayton}}, \bibinfo
  {author} {\bibfnamefont {J.-R.}\ \bibnamefont {Gao}}, \bibinfo {author}
  {\bibfnamefont {J.~L.}\ \bibnamefont {Reno}},\ and\ \bibinfo {author}
  {\bibfnamefont {Q.}~\bibnamefont {Hu}},\ }\bibfield  {title}
  {{\selectlanguage {en}\bibinfo {title} {Terahertz multiheterodyne
  spectroscopy using laser frequency combs}},\ }\href
  {https://doi.org/10.1364/OPTICA.3.000499} {\bibfield  {journal} {\bibinfo
  {journal} {Optica}\ }\textbf {\bibinfo {volume} {3}},\ \bibinfo {pages} {499}
  (\bibinfo {year} {2016})}\BibitemShut {NoStop}%
\bibitem [{\citenamefont {Yu}\ \emph {et~al.}(2018)\citenamefont {Yu},
  \citenamefont {Okawachi}, \citenamefont {Griffith}, \citenamefont
  {Picqu{\'e}}, \citenamefont {Lipson},\ and\ \citenamefont
  {Gaeta}}]{yuSiliconchipbasedMidinfraredDualcomb2018}%
  \BibitemOpen
  \bibfield  {author} {\bibinfo {author} {\bibfnamefont {M.}~\bibnamefont
  {Yu}}, \bibinfo {author} {\bibfnamefont {Y.}~\bibnamefont {Okawachi}},
  \bibinfo {author} {\bibfnamefont {A.~G.}\ \bibnamefont {Griffith}}, \bibinfo
  {author} {\bibfnamefont {N.}~\bibnamefont {Picqu{\'e}}}, \bibinfo {author}
  {\bibfnamefont {M.}~\bibnamefont {Lipson}},\ and\ \bibinfo {author}
  {\bibfnamefont {A.~L.}\ \bibnamefont {Gaeta}},\ }\bibfield  {title}
  {{\selectlanguage {en}\bibinfo {title} {Silicon-chip-based mid-infrared
  dual-comb spectroscopy}},\ }\href
  {https://doi.org/10.1038/s41467-018-04350-1} {\bibfield  {journal} {\bibinfo
  {journal} {Nature Communications}\ }\textbf {\bibinfo {volume} {9}},\
  \bibinfo {pages} {1869} (\bibinfo {year} {2018})}\BibitemShut {NoStop}%
\bibitem [{\citenamefont {Suh}\ \emph {et~al.}(2016)\citenamefont {Suh},
  \citenamefont {Yang}, \citenamefont {Yang}, \citenamefont {Yi},\ and\
  \citenamefont {Vahala}}]{suhMicroresonatorSolitonDualcomb2016}%
  \BibitemOpen
  \bibfield  {author} {\bibinfo {author} {\bibfnamefont {M.-G.}\ \bibnamefont
  {Suh}}, \bibinfo {author} {\bibfnamefont {Q.-F.}\ \bibnamefont {Yang}},
  \bibinfo {author} {\bibfnamefont {K.~Y.}\ \bibnamefont {Yang}}, \bibinfo
  {author} {\bibfnamefont {X.}~\bibnamefont {Yi}},\ and\ \bibinfo {author}
  {\bibfnamefont {K.~J.}\ \bibnamefont {Vahala}},\ }\bibfield  {title}
  {{\selectlanguage {en}\bibinfo {title} {Microresonator soliton dual-comb
  spectroscopy}},\ }\href {https://doi.org/10.1126/science.aah6516} {\bibfield
  {journal} {\bibinfo  {journal} {Science}\ }\textbf {\bibinfo {volume}
  {354}},\ \bibinfo {pages} {600} (\bibinfo {year} {2016})}\BibitemShut
  {NoStop}%
\bibitem [{\citenamefont {Del'Haye}\ \emph {et~al.}(2009)\citenamefont
  {Del'Haye}, \citenamefont {Arcizet}, \citenamefont {Gorodetsky},
  \citenamefont {Holzwarth},\ and\ \citenamefont
  {Kippenberg}}]{delhayeFrequencyCombAssisted2009}%
  \BibitemOpen
  \bibfield  {author} {\bibinfo {author} {\bibfnamefont {P.}~\bibnamefont
  {Del'Haye}}, \bibinfo {author} {\bibfnamefont {O.}~\bibnamefont {Arcizet}},
  \bibinfo {author} {\bibfnamefont {M.~L.}\ \bibnamefont {Gorodetsky}},
  \bibinfo {author} {\bibfnamefont {R.}~\bibnamefont {Holzwarth}},\ and\
  \bibinfo {author} {\bibfnamefont {T.~J.}\ \bibnamefont {Kippenberg}},\
  }\bibfield  {title} {{\selectlanguage {en}\bibinfo {title} {Frequency comb
  assisted diode laser spectroscopy for measurement of microcavity
  dispersion}},\ }\href {https://doi.org/10.1038/nphoton.2009.138} {\bibfield
  {journal} {\bibinfo  {journal} {Nature Photonics}\ }\textbf {\bibinfo
  {volume} {3}},\ \bibinfo {pages} {529} (\bibinfo {year} {2009})}\BibitemShut
  {NoStop}%
\bibitem [{\citenamefont {Bartalini}\ \emph {et~al.}(2014)\citenamefont
  {Bartalini}, \citenamefont {Consolino}, \citenamefont {Cancio}, \citenamefont
  {De~Natale}, \citenamefont {Bartolini}, \citenamefont {Taschin},
  \citenamefont {De~Pas}, \citenamefont {Beere}, \citenamefont {Ritchie},
  \citenamefont {Vitiello},\ and\ \citenamefont
  {Torre}}]{bartaliniFrequencyCombAssistedTerahertzQuantum2014}%
  \BibitemOpen
  \bibfield  {author} {\bibinfo {author} {\bibfnamefont {S.}~\bibnamefont
  {Bartalini}}, \bibinfo {author} {\bibfnamefont {L.}~\bibnamefont
  {Consolino}}, \bibinfo {author} {\bibfnamefont {P.}~\bibnamefont {Cancio}},
  \bibinfo {author} {\bibfnamefont {P.}~\bibnamefont {De~Natale}}, \bibinfo
  {author} {\bibfnamefont {P.}~\bibnamefont {Bartolini}}, \bibinfo {author}
  {\bibfnamefont {A.}~\bibnamefont {Taschin}}, \bibinfo {author} {\bibfnamefont
  {M.}~\bibnamefont {De~Pas}}, \bibinfo {author} {\bibfnamefont
  {H.}~\bibnamefont {Beere}}, \bibinfo {author} {\bibfnamefont
  {D.}~\bibnamefont {Ritchie}}, \bibinfo {author} {\bibfnamefont {M.~S.}\
  \bibnamefont {Vitiello}},\ and\ \bibinfo {author} {\bibfnamefont
  {R.}~\bibnamefont {Torre}},\ }\bibfield  {title} {{\selectlanguage
  {en}\bibinfo {title} {Frequency-{{Comb}}-{{Assisted Terahertz Quantum Cascade
  Laser Spectroscopy}}}},\ }\href {https://doi.org/10.1103/PhysRevX.4.021006}
  {\bibfield  {journal} {\bibinfo  {journal} {Physical Review X}\ }\textbf
  {\bibinfo {volume} {4}},\ \bibinfo {pages} {021006} (\bibinfo {year}
  {2014})}\BibitemShut {NoStop}%
\bibitem [{\citenamefont {Giorgetta}\ \emph {et~al.}(2010)\citenamefont
  {Giorgetta}, \citenamefont {Coddington}, \citenamefont {Baumann},
  \citenamefont {Swann},\ and\ \citenamefont
  {Newbury}}]{giorgettaFastHighresolutionSpectroscopy2010}%
  \BibitemOpen
  \bibfield  {author} {\bibinfo {author} {\bibfnamefont {F.~R.}\ \bibnamefont
  {Giorgetta}}, \bibinfo {author} {\bibfnamefont {I.}~\bibnamefont
  {Coddington}}, \bibinfo {author} {\bibfnamefont {E.}~\bibnamefont {Baumann}},
  \bibinfo {author} {\bibfnamefont {W.~C.}\ \bibnamefont {Swann}},\ and\
  \bibinfo {author} {\bibfnamefont {N.~R.}\ \bibnamefont {Newbury}},\
  }\bibfield  {title} {{\selectlanguage {en}\bibinfo {title} {Fast
  high-resolution spectroscopy of dynamic continuous-wave laser sources}},\
  }\href {https://doi.org/10.1038/nphoton.2010.228} {\bibfield  {journal}
  {\bibinfo  {journal} {Nature Photonics}\ }\textbf {\bibinfo {volume} {4}},\
  \bibinfo {pages} {853} (\bibinfo {year} {2010})}\BibitemShut {NoStop}%
\bibitem [{\citenamefont {Coddington}\ \emph {et~al.}(2012)\citenamefont
  {Coddington}, \citenamefont {Giorgetta}, \citenamefont {Baumann},
  \citenamefont {Swann},\ and\ \citenamefont
  {Newbury}}]{coddingtonCharacterizingFastArbitrary2012}%
  \BibitemOpen
  \bibfield  {author} {\bibinfo {author} {\bibfnamefont {I.}~\bibnamefont
  {Coddington}}, \bibinfo {author} {\bibfnamefont {F.~R.}\ \bibnamefont
  {Giorgetta}}, \bibinfo {author} {\bibfnamefont {E.}~\bibnamefont {Baumann}},
  \bibinfo {author} {\bibfnamefont {W.~C.}\ \bibnamefont {Swann}},\ and\
  \bibinfo {author} {\bibfnamefont {N.~R.}\ \bibnamefont {Newbury}},\
  }\bibfield  {title} {\bibinfo {title} {Characterizing {{Fast Arbitrary CW
  Waveforms With}} 1500 {{THz}}/s {{Instantaneous Chirps}}},\ }\href
  {https://doi.org/10.1109/JSTQE.2011.2114875} {\bibfield  {journal} {\bibinfo
  {journal} {IEEE Journal of Selected Topics in Quantum Electronics}\ }\textbf
  {\bibinfo {volume} {18}},\ \bibinfo {pages} {228} (\bibinfo {year}
  {2012})}\BibitemShut {NoStop}%
\bibitem [{\citenamefont {Yang}\ \emph {et~al.}(2019)\citenamefont {Yang},
  \citenamefont {Shen}, \citenamefont {Wang}, \citenamefont {Tran},
  \citenamefont {Zhang}, \citenamefont {Yang}, \citenamefont {Wu},
  \citenamefont {Bao}, \citenamefont {Bowers}, \citenamefont {Yariv},\ and\
  \citenamefont {Vahala}}]{yangVernierSpectrometerUsing2019}%
  \BibitemOpen
  \bibfield  {author} {\bibinfo {author} {\bibfnamefont {Q.-F.}\ \bibnamefont
  {Yang}}, \bibinfo {author} {\bibfnamefont {B.}~\bibnamefont {Shen}}, \bibinfo
  {author} {\bibfnamefont {H.}~\bibnamefont {Wang}}, \bibinfo {author}
  {\bibfnamefont {M.}~\bibnamefont {Tran}}, \bibinfo {author} {\bibfnamefont
  {Z.}~\bibnamefont {Zhang}}, \bibinfo {author} {\bibfnamefont {K.~Y.}\
  \bibnamefont {Yang}}, \bibinfo {author} {\bibfnamefont {L.}~\bibnamefont
  {Wu}}, \bibinfo {author} {\bibfnamefont {C.}~\bibnamefont {Bao}}, \bibinfo
  {author} {\bibfnamefont {J.}~\bibnamefont {Bowers}}, \bibinfo {author}
  {\bibfnamefont {A.}~\bibnamefont {Yariv}},\ and\ \bibinfo {author}
  {\bibfnamefont {K.}~\bibnamefont {Vahala}},\ }\bibfield  {title}
  {{\selectlanguage {en}\bibinfo {title} {Vernier spectrometer using
  counterpropagating soliton microcombs}},\ }\href
  {https://doi.org/10.1126/science.aaw2317} {\bibfield  {journal} {\bibinfo
  {journal} {Science}\ }\textbf {\bibinfo {volume} {363}},\ \bibinfo {pages}
  {965} (\bibinfo {year} {2019})}\BibitemShut {NoStop}%
\bibitem [{\citenamefont {Mandon}\ \emph {et~al.}(2009)\citenamefont {Mandon},
  \citenamefont {Guelachvili},\ and\ \citenamefont
  {Picqu{\'e}}}]{mandonFourierTransformSpectroscopy2009}%
  \BibitemOpen
  \bibfield  {author} {\bibinfo {author} {\bibfnamefont {J.}~\bibnamefont
  {Mandon}}, \bibinfo {author} {\bibfnamefont {G.}~\bibnamefont
  {Guelachvili}},\ and\ \bibinfo {author} {\bibfnamefont {N.}~\bibnamefont
  {Picqu{\'e}}},\ }\bibfield  {title} {{\selectlanguage {en}\bibinfo {title}
  {Fourier transform spectroscopy with a laser frequency comb}},\ }\href
  {https://doi.org/10.1038/nphoton.2008.293} {\bibfield  {journal} {\bibinfo
  {journal} {Nature Photonics}\ }\textbf {\bibinfo {volume} {3}},\ \bibinfo
  {pages} {99} (\bibinfo {year} {2009})}\BibitemShut {NoStop}%
\bibitem [{\citenamefont {Hugi}\ \emph {et~al.}(2012)\citenamefont {Hugi},
  \citenamefont {Villares}, \citenamefont {Blaser}, \citenamefont {Liu},\ and\
  \citenamefont {Faist}}]{hugiMidinfraredFrequencyComb2012}%
  \BibitemOpen
  \bibfield  {author} {\bibinfo {author} {\bibfnamefont {A.}~\bibnamefont
  {Hugi}}, \bibinfo {author} {\bibfnamefont {G.}~\bibnamefont {Villares}},
  \bibinfo {author} {\bibfnamefont {S.}~\bibnamefont {Blaser}}, \bibinfo
  {author} {\bibfnamefont {H.~C.}\ \bibnamefont {Liu}},\ and\ \bibinfo {author}
  {\bibfnamefont {J.}~\bibnamefont {Faist}},\ }\bibfield  {title}
  {{\selectlanguage {en}\bibinfo {title} {Mid-infrared frequency comb based on
  a quantum cascade laser}},\ }\href {https://doi.org/10.1038/nature11620}
  {\bibfield  {journal} {\bibinfo  {journal} {Nature}\ }\textbf {\bibinfo
  {volume} {492}},\ \bibinfo {pages} {229} (\bibinfo {year}
  {2012})}\BibitemShut {NoStop}%
\bibitem [{\citenamefont {Burghoff}\ \emph {et~al.}(2014)\citenamefont
  {Burghoff}, \citenamefont {Kao}, \citenamefont {Han}, \citenamefont {Chan},
  \citenamefont {Cai}, \citenamefont {Yang}, \citenamefont {Hayton},
  \citenamefont {Gao}, \citenamefont {Reno},\ and\ \citenamefont
  {Hu}}]{burghoffTerahertzLaserFrequency2014}%
  \BibitemOpen
  \bibfield  {author} {\bibinfo {author} {\bibfnamefont {D.}~\bibnamefont
  {Burghoff}}, \bibinfo {author} {\bibfnamefont {T.-Y.}\ \bibnamefont {Kao}},
  \bibinfo {author} {\bibfnamefont {N.}~\bibnamefont {Han}}, \bibinfo {author}
  {\bibfnamefont {C.~W.~I.}\ \bibnamefont {Chan}}, \bibinfo {author}
  {\bibfnamefont {X.}~\bibnamefont {Cai}}, \bibinfo {author} {\bibfnamefont
  {Y.}~\bibnamefont {Yang}}, \bibinfo {author} {\bibfnamefont {D.~J.}\
  \bibnamefont {Hayton}}, \bibinfo {author} {\bibfnamefont {J.-R.}\
  \bibnamefont {Gao}}, \bibinfo {author} {\bibfnamefont {J.~L.}\ \bibnamefont
  {Reno}},\ and\ \bibinfo {author} {\bibfnamefont {Q.}~\bibnamefont {Hu}},\
  }\bibfield  {title} {{\selectlanguage {en}\bibinfo {title} {Terahertz laser
  frequency combs}},\ }\href {https://doi.org/10.1038/nphoton.2014.85}
  {\bibfield  {journal} {\bibinfo  {journal} {Nature Photonics}\ }\textbf
  {\bibinfo {volume} {8}},\ \bibinfo {pages} {462} (\bibinfo {year}
  {2014})}\BibitemShut {NoStop}%
\bibitem [{\citenamefont {Burghoff}\ \emph {et~al.}(2015)\citenamefont
  {Burghoff}, \citenamefont {Yang}, \citenamefont {Hayton}, \citenamefont
  {Gao}, \citenamefont {Reno},\ and\ \citenamefont
  {Hu}}]{burghoffEvaluatingCoherenceTimedomain2015}%
  \BibitemOpen
  \bibfield  {author} {\bibinfo {author} {\bibfnamefont {D.}~\bibnamefont
  {Burghoff}}, \bibinfo {author} {\bibfnamefont {Y.}~\bibnamefont {Yang}},
  \bibinfo {author} {\bibfnamefont {D.~J.}\ \bibnamefont {Hayton}}, \bibinfo
  {author} {\bibfnamefont {J.-R.}\ \bibnamefont {Gao}}, \bibinfo {author}
  {\bibfnamefont {J.~L.}\ \bibnamefont {Reno}},\ and\ \bibinfo {author}
  {\bibfnamefont {Q.}~\bibnamefont {Hu}},\ }\bibfield  {title}
  {{\selectlanguage {en}\bibinfo {title} {Evaluating the coherence and
  time-domain profile of quantum cascade laser frequency combs}},\ }\href
  {https://doi.org/10.1364/OE.23.001190} {\bibfield  {journal} {\bibinfo
  {journal} {Optics Express}\ }\textbf {\bibinfo {volume} {23}},\ \bibinfo
  {pages} {1190} (\bibinfo {year} {2015})}\BibitemShut {NoStop}%
\bibitem [{\citenamefont {Han}\ \emph {et~al.}(2020)\citenamefont {Han},
  \citenamefont {Ren},\ and\ \citenamefont
  {Burghoff}}]{hanSensitivitySWIFTSpectroscopy2020}%
  \BibitemOpen
  \bibfield  {author} {\bibinfo {author} {\bibfnamefont {Z.}~\bibnamefont
  {Han}}, \bibinfo {author} {\bibfnamefont {D.}~\bibnamefont {Ren}},\ and\
  \bibinfo {author} {\bibfnamefont {D.}~\bibnamefont {Burghoff}},\ }\bibfield
  {title} {{\selectlanguage {EN}\bibinfo {title} {Sensitivity of {{SWIFT}}
  spectroscopy}},\ }\href {https://doi.org/10.1364/OE.382243} {\bibfield
  {journal} {\bibinfo  {journal} {Optics Express}\ }\textbf {\bibinfo {volume}
  {28}},\ \bibinfo {pages} {6002} (\bibinfo {year} {2020})}\BibitemShut
  {NoStop}%
\bibitem [{\citenamefont
  {Fellgett}(1951)}]{fellgettTheoryInfraredSensitivities1951}%
  \BibitemOpen
  \bibfield  {author} {\bibinfo {author} {\bibfnamefont {P.~B.}\ \bibnamefont
  {Fellgett}},\ }\href@noop {} {{\selectlanguage {en}\emph {\bibinfo {title}
  {The {{Theory}} of {{Infrared Sensitivities}} and {{Its Application}} to
  {{Investigations}} of {{Stellar Radiation}} in the {{Near Infra}}-Red}}}}\
  (\bibinfo {year} {1951})\BibitemShut {NoStop}%
\bibitem [{\citenamefont
  {Jacquinot}(1958)}]{jacquinotCaracteresCommunsAux1958}%
  \BibitemOpen
  \bibfield  {author} {\bibinfo {author} {\bibfnamefont {P.}~\bibnamefont
  {Jacquinot}},\ }\bibfield  {title} {{\selectlanguage {fr}\bibinfo {title}
  {{Caract{\`e}res communs aux nouvelles m{\'e}thodes de spectroscopie
  interf{\'e}rentielle ; Facteur de m{\'e}rite}}},\ }\href
  {https://doi.org/10.1051/jphysrad:01958001903022300} {\bibfield  {journal}
  {\bibinfo  {journal} {Journal de Physique et le Radium}\ }\textbf {\bibinfo
  {volume} {19}},\ \bibinfo {pages} {223} (\bibinfo {year} {1958})}\BibitemShut
  {NoStop}%
\bibitem [{\citenamefont {Hillbrand}\ \emph {et~al.}(2020)\citenamefont
  {Hillbrand}, \citenamefont {Auth}, \citenamefont {Piccardo}, \citenamefont
  {Opa{\v c}ak}, \citenamefont {Gornik}, \citenamefont {Strasser},
  \citenamefont {Capasso}, \citenamefont {Breuer},\ and\ \citenamefont
  {Schwarz}}]{hillbrandInPhaseAntiPhaseSynchronization2020}%
  \BibitemOpen
  \bibfield  {author} {\bibinfo {author} {\bibfnamefont {J.}~\bibnamefont
  {Hillbrand}}, \bibinfo {author} {\bibfnamefont {D.}~\bibnamefont {Auth}},
  \bibinfo {author} {\bibfnamefont {M.}~\bibnamefont {Piccardo}}, \bibinfo
  {author} {\bibfnamefont {N.}~\bibnamefont {Opa{\v c}ak}}, \bibinfo {author}
  {\bibfnamefont {E.}~\bibnamefont {Gornik}}, \bibinfo {author} {\bibfnamefont
  {G.}~\bibnamefont {Strasser}}, \bibinfo {author} {\bibfnamefont
  {F.}~\bibnamefont {Capasso}}, \bibinfo {author} {\bibfnamefont
  {S.}~\bibnamefont {Breuer}},\ and\ \bibinfo {author} {\bibfnamefont
  {B.}~\bibnamefont {Schwarz}},\ }\bibfield  {title} {\bibinfo {title}
  {In-{{Phase}} and {{Anti}}-{{Phase Synchronization}} in a {{Laser Frequency
  Comb}}},\ }\href {https://doi.org/10.1103/PhysRevLett.124.023901} {\bibfield
  {journal} {\bibinfo  {journal} {Physical Review Letters}\ }\textbf {\bibinfo
  {volume} {124}},\ \bibinfo {pages} {023901} (\bibinfo {year}
  {2020})}\BibitemShut {NoStop}%
\bibitem [{\citenamefont {Bagheri}\ \emph {et~al.}(2018)\citenamefont
  {Bagheri}, \citenamefont {Frez}, \citenamefont {Sterczewski}, \citenamefont
  {Gruidin}, \citenamefont {Fradet}, \citenamefont {Vurgaftman}, \citenamefont
  {Canedy}, \citenamefont {Bewley}, \citenamefont {Merritt}, \citenamefont
  {Kim}, \citenamefont {Kim},\ and\ \citenamefont
  {Meyer}}]{bagheriPassivelyModelockedInterband2018}%
  \BibitemOpen
  \bibfield  {author} {\bibinfo {author} {\bibfnamefont {M.}~\bibnamefont
  {Bagheri}}, \bibinfo {author} {\bibfnamefont {C.}~\bibnamefont {Frez}},
  \bibinfo {author} {\bibfnamefont {L.~A.}\ \bibnamefont {Sterczewski}},
  \bibinfo {author} {\bibfnamefont {I.}~\bibnamefont {Gruidin}}, \bibinfo
  {author} {\bibfnamefont {M.}~\bibnamefont {Fradet}}, \bibinfo {author}
  {\bibfnamefont {I.}~\bibnamefont {Vurgaftman}}, \bibinfo {author}
  {\bibfnamefont {C.~L.}\ \bibnamefont {Canedy}}, \bibinfo {author}
  {\bibfnamefont {W.~W.}\ \bibnamefont {Bewley}}, \bibinfo {author}
  {\bibfnamefont {C.~D.}\ \bibnamefont {Merritt}}, \bibinfo {author}
  {\bibfnamefont {C.~S.}\ \bibnamefont {Kim}}, \bibinfo {author} {\bibfnamefont
  {M.}~\bibnamefont {Kim}},\ and\ \bibinfo {author} {\bibfnamefont {J.~R.}\
  \bibnamefont {Meyer}},\ }\bibfield  {title} {{\selectlanguage {en}\bibinfo
  {title} {Passively mode-locked interband cascade optical frequency combs}},\
  }\href {https://doi.org/10.1038/s41598-018-21504-9} {\bibfield  {journal}
  {\bibinfo  {journal} {Scientific Reports}\ }\textbf {\bibinfo {volume} {8}},\
  \bibinfo {pages} {1} (\bibinfo {year} {2018})}\BibitemShut {NoStop}%
\bibitem [{\citenamefont {Singleton}\ \emph {et~al.}(2018)\citenamefont
  {Singleton}, \citenamefont {Jouy}, \citenamefont {Beck},\ and\ \citenamefont
  {Faist}}]{singletonEvidenceLinearChirp2018}%
  \BibitemOpen
  \bibfield  {author} {\bibinfo {author} {\bibfnamefont {M.}~\bibnamefont
  {Singleton}}, \bibinfo {author} {\bibfnamefont {P.}~\bibnamefont {Jouy}},
  \bibinfo {author} {\bibfnamefont {M.}~\bibnamefont {Beck}},\ and\ \bibinfo
  {author} {\bibfnamefont {J.}~\bibnamefont {Faist}},\ }\bibfield  {title}
  {{\selectlanguage {en}\bibinfo {title} {Evidence of linear chirp in
  mid-infrared quantum cascade lasers}},\ }\href
  {https://doi.org/10.1364/OPTICA.5.000948} {\bibfield  {journal} {\bibinfo
  {journal} {Optica}\ }\textbf {\bibinfo {volume} {5}},\ \bibinfo {pages} {948}
  (\bibinfo {year} {2018})}\BibitemShut {NoStop}%
\bibitem [{\citenamefont {Dong}\ \emph {et~al.}(2018)\citenamefont {Dong},
  \citenamefont {Cundiff},\ and\ \citenamefont
  {Winful}}]{dongPhysicsFrequencymodulatedComb2018}%
  \BibitemOpen
  \bibfield  {author} {\bibinfo {author} {\bibfnamefont {M.}~\bibnamefont
  {Dong}}, \bibinfo {author} {\bibfnamefont {S.~T.}\ \bibnamefont {Cundiff}},\
  and\ \bibinfo {author} {\bibfnamefont {H.~G.}\ \bibnamefont {Winful}},\
  }\bibfield  {title} {\bibinfo {title} {Physics of frequency-modulated comb
  generation in quantum-well diode lasers},\ }\href
  {https://doi.org/10.1103/PhysRevA.97.053822} {\bibfield  {journal} {\bibinfo
  {journal} {Physical Review A}\ }\textbf {\bibinfo {volume} {97}},\ \bibinfo
  {pages} {053822} (\bibinfo {year} {2018})}\BibitemShut {NoStop}%
\bibitem [{\citenamefont {Henry}\ \emph {et~al.}(2017)\citenamefont {Henry},
  \citenamefont {Burghoff}, \citenamefont {Yang}, \citenamefont {Hu},\ and\
  \citenamefont {Khurgin}}]{henryPseudorandomDynamicsFrequency2017}%
  \BibitemOpen
  \bibfield  {author} {\bibinfo {author} {\bibfnamefont {N.}~\bibnamefont
  {Henry}}, \bibinfo {author} {\bibfnamefont {D.}~\bibnamefont {Burghoff}},
  \bibinfo {author} {\bibfnamefont {Y.}~\bibnamefont {Yang}}, \bibinfo {author}
  {\bibfnamefont {Q.}~\bibnamefont {Hu}},\ and\ \bibinfo {author}
  {\bibfnamefont {J.~B.}\ \bibnamefont {Khurgin}},\ }\bibfield  {title}
  {{\selectlanguage {en}\bibinfo {title} {Pseudorandom dynamics of frequency
  combs in free-running quantum cascade lasers}},\ }\href
  {https://doi.org/10.1117/1.OE.57.1.011009} {\bibfield  {journal} {\bibinfo
  {journal} {Optical Engineering}\ }\textbf {\bibinfo {volume} {57}},\ \bibinfo
  {pages} {1} (\bibinfo {year} {2017})}\BibitemShut {NoStop}%
\bibitem [{\citenamefont {Cappelli}\ \emph {et~al.}(2019)\citenamefont
  {Cappelli}, \citenamefont {Consolino}, \citenamefont {Campo}, \citenamefont
  {Galli}, \citenamefont {Mazzotti}, \citenamefont {Campa}, \citenamefont
  {de~Cumis}, \citenamefont {Pastor}, \citenamefont {Eramo}, \citenamefont
  {R{\"o}sch}, \citenamefont {Beck}, \citenamefont {Scalari}, \citenamefont
  {Faist}, \citenamefont {Natale},\ and\ \citenamefont
  {Bartalini}}]{cappelliRetrievalPhaseRelation2019}%
  \BibitemOpen
  \bibfield  {author} {\bibinfo {author} {\bibfnamefont {F.}~\bibnamefont
  {Cappelli}}, \bibinfo {author} {\bibfnamefont {L.}~\bibnamefont {Consolino}},
  \bibinfo {author} {\bibfnamefont {G.}~\bibnamefont {Campo}}, \bibinfo
  {author} {\bibfnamefont {I.}~\bibnamefont {Galli}}, \bibinfo {author}
  {\bibfnamefont {D.}~\bibnamefont {Mazzotti}}, \bibinfo {author}
  {\bibfnamefont {A.}~\bibnamefont {Campa}}, \bibinfo {author} {\bibfnamefont
  {M.~S.}\ \bibnamefont {de~Cumis}}, \bibinfo {author} {\bibfnamefont {P.~C.}\
  \bibnamefont {Pastor}}, \bibinfo {author} {\bibfnamefont {R.}~\bibnamefont
  {Eramo}}, \bibinfo {author} {\bibfnamefont {M.}~\bibnamefont {R{\"o}sch}},
  \bibinfo {author} {\bibfnamefont {M.}~\bibnamefont {Beck}}, \bibinfo {author}
  {\bibfnamefont {G.}~\bibnamefont {Scalari}}, \bibinfo {author} {\bibfnamefont
  {J.}~\bibnamefont {Faist}}, \bibinfo {author} {\bibfnamefont {P.~D.}\
  \bibnamefont {Natale}},\ and\ \bibinfo {author} {\bibfnamefont
  {S.}~\bibnamefont {Bartalini}},\ }\bibfield  {title} {{\selectlanguage
  {en}\bibinfo {title} {Retrieval of phase relation and emission profile of
  quantum cascade laser frequency combs}},\ }\href
  {https://doi.org/10.1038/s41566-019-0451-1} {\bibfield  {journal} {\bibinfo
  {journal} {Nature Photonics}\ }\textbf {\bibinfo {volume} {13}},\ \bibinfo
  {pages} {562} (\bibinfo {year} {2019})}\BibitemShut {NoStop}%
\bibitem [{\citenamefont {Del'Haye}\ \emph {et~al.}(2015)\citenamefont
  {Del'Haye}, \citenamefont {Coillet}, \citenamefont {Loh}, \citenamefont
  {Beha}, \citenamefont {Papp},\ and\ \citenamefont
  {Diddams}}]{delhayePhaseStepsResonator2015}%
  \BibitemOpen
  \bibfield  {author} {\bibinfo {author} {\bibfnamefont {P.}~\bibnamefont
  {Del'Haye}}, \bibinfo {author} {\bibfnamefont {A.}~\bibnamefont {Coillet}},
  \bibinfo {author} {\bibfnamefont {W.}~\bibnamefont {Loh}}, \bibinfo {author}
  {\bibfnamefont {K.}~\bibnamefont {Beha}}, \bibinfo {author} {\bibfnamefont
  {S.~B.}\ \bibnamefont {Papp}},\ and\ \bibinfo {author} {\bibfnamefont
  {S.~A.}\ \bibnamefont {Diddams}},\ }\bibfield  {title} {{\selectlanguage
  {en}\bibinfo {title} {Phase steps and resonator detuning measurements in
  microresonator frequency combs}},\ }\href
  {https://doi.org/10.1038/ncomms6668} {\bibfield  {journal} {\bibinfo
  {journal} {Nature Communications}\ }\textbf {\bibinfo {volume} {6}},\
  \bibinfo {pages} {5668} (\bibinfo {year} {2015})}\BibitemShut {NoStop}%
\bibitem [{\citenamefont {Metcalf}\ \emph {et~al.}(2013)\citenamefont
  {Metcalf}, \citenamefont {{Torres-Company}}, \citenamefont {Leaird},\ and\
  \citenamefont {Weiner}}]{metcalfHighPowerBroadlyTunable2013}%
  \BibitemOpen
  \bibfield  {author} {\bibinfo {author} {\bibfnamefont {A.~J.}\ \bibnamefont
  {Metcalf}}, \bibinfo {author} {\bibfnamefont {V.}~\bibnamefont
  {{Torres-Company}}}, \bibinfo {author} {\bibfnamefont {D.~E.}\ \bibnamefont
  {Leaird}},\ and\ \bibinfo {author} {\bibfnamefont {A.~M.}\ \bibnamefont
  {Weiner}},\ }\bibfield  {title} {\bibinfo {title} {High-{{Power Broadly
  Tunable Electrooptic Frequency Comb Generator}}},\ }\href
  {https://doi.org/10.1109/JSTQE.2013.2268384} {\bibfield  {journal} {\bibinfo
  {journal} {IEEE Journal of Selected Topics in Quantum Electronics}\ }\textbf
  {\bibinfo {volume} {19}},\ \bibinfo {pages} {231} (\bibinfo {year}
  {2013})}\BibitemShut {NoStop}%
\bibitem [{\citenamefont {Mashian}\ \emph {et~al.}(2015)\citenamefont
  {Mashian}, \citenamefont {Sturm}, \citenamefont {Sternberg}, \citenamefont
  {Janssen}, \citenamefont {{Hailey-Dunsheath}}, \citenamefont {Fischer},
  \citenamefont {Contursi}, \citenamefont {{Gonz{\'a}lez-Alfonso}},
  \citenamefont {{Graci{\'a}-Carpio}}, \citenamefont {Poglitsch}, \citenamefont
  {Veilleux}, \citenamefont {Davies}, \citenamefont {Genzel}, \citenamefont
  {Lutz}, \citenamefont {Tacconi}, \citenamefont {Verma}, \citenamefont
  {Weiss}, \citenamefont {Polisensky},\ and\ \citenamefont
  {Nikola}}]{mashianHighJCOSLEDs2015}%
  \BibitemOpen
  \bibfield  {author} {\bibinfo {author} {\bibfnamefont {N.}~\bibnamefont
  {Mashian}}, \bibinfo {author} {\bibfnamefont {E.}~\bibnamefont {Sturm}},
  \bibinfo {author} {\bibfnamefont {A.}~\bibnamefont {Sternberg}}, \bibinfo
  {author} {\bibfnamefont {A.}~\bibnamefont {Janssen}}, \bibinfo {author}
  {\bibfnamefont {S.}~\bibnamefont {{Hailey-Dunsheath}}}, \bibinfo {author}
  {\bibfnamefont {J.}~\bibnamefont {Fischer}}, \bibinfo {author} {\bibfnamefont
  {A.}~\bibnamefont {Contursi}}, \bibinfo {author} {\bibfnamefont
  {E.}~\bibnamefont {{Gonz{\'a}lez-Alfonso}}}, \bibinfo {author} {\bibfnamefont
  {J.}~\bibnamefont {{Graci{\'a}-Carpio}}}, \bibinfo {author} {\bibfnamefont
  {A.}~\bibnamefont {Poglitsch}}, \bibinfo {author} {\bibfnamefont
  {S.}~\bibnamefont {Veilleux}}, \bibinfo {author} {\bibfnamefont
  {R.}~\bibnamefont {Davies}}, \bibinfo {author} {\bibfnamefont
  {R.}~\bibnamefont {Genzel}}, \bibinfo {author} {\bibfnamefont
  {D.}~\bibnamefont {Lutz}}, \bibinfo {author} {\bibfnamefont {L.}~\bibnamefont
  {Tacconi}}, \bibinfo {author} {\bibfnamefont {A.}~\bibnamefont {Verma}},
  \bibinfo {author} {\bibfnamefont {A.}~\bibnamefont {Weiss}}, \bibinfo
  {author} {\bibfnamefont {E.}~\bibnamefont {Polisensky}},\ and\ \bibinfo
  {author} {\bibfnamefont {T.}~\bibnamefont {Nikola}},\ }\bibfield  {title}
  {{\selectlanguage {en}\bibinfo {title} {High-{{J CO SLEDs}} in nearby
  infrared bright galaxies observed by {{Herschel}}/{{PACS}}}},\ }\href
  {https://doi.org/10.1088/0004-637X/802/2/81} {\bibfield  {journal} {\bibinfo
  {journal} {The Astrophysical Journal}\ }\textbf {\bibinfo {volume} {802}},\
  \bibinfo {pages} {81} (\bibinfo {year} {2015})}\BibitemShut {NoStop}%
\bibitem [{\citenamefont {Faist}\ \emph {et~al.}(2016)\citenamefont {Faist},
  \citenamefont {Villares}, \citenamefont {Scalari}, \citenamefont {R{\"o}sch},
  \citenamefont {Bonzon}, \citenamefont {Hugi},\ and\ \citenamefont
  {Beck}}]{faistQuantumCascadeLaser2016}%
  \BibitemOpen
  \bibfield  {author} {\bibinfo {author} {\bibfnamefont {J.}~\bibnamefont
  {Faist}}, \bibinfo {author} {\bibfnamefont {G.}~\bibnamefont {Villares}},
  \bibinfo {author} {\bibfnamefont {G.}~\bibnamefont {Scalari}}, \bibinfo
  {author} {\bibfnamefont {M.}~\bibnamefont {R{\"o}sch}}, \bibinfo {author}
  {\bibfnamefont {C.}~\bibnamefont {Bonzon}}, \bibinfo {author} {\bibfnamefont
  {A.}~\bibnamefont {Hugi}},\ and\ \bibinfo {author} {\bibfnamefont
  {M.}~\bibnamefont {Beck}},\ }\bibfield  {title} {{\selectlanguage
  {en}\bibinfo {title} {Quantum {{Cascade Laser Frequency Combs}}}},\
  }\bibfield  {journal} {\bibinfo  {journal} {Nanophotonics}\ }\textbf
  {\bibinfo {volume} {5}},\ \href {https://doi.org/10.1515/nanoph-2016-0015}
  {10.1515/nanoph-2016-0015} (\bibinfo {year} {2016})\BibitemShut {NoStop}%
\bibitem [{\citenamefont {Ideguchi}\ \emph {et~al.}(2014)\citenamefont
  {Ideguchi}, \citenamefont {Poisson}, \citenamefont {Guelachvili},
  \citenamefont {Picqu{\'e}},\ and\ \citenamefont
  {H{\"a}nsch}}]{ideguchiAdaptiveRealtimeDualcomb2014}%
  \BibitemOpen
  \bibfield  {author} {\bibinfo {author} {\bibfnamefont {T.}~\bibnamefont
  {Ideguchi}}, \bibinfo {author} {\bibfnamefont {A.}~\bibnamefont {Poisson}},
  \bibinfo {author} {\bibfnamefont {G.}~\bibnamefont {Guelachvili}}, \bibinfo
  {author} {\bibfnamefont {N.}~\bibnamefont {Picqu{\'e}}},\ and\ \bibinfo
  {author} {\bibfnamefont {T.~W.}\ \bibnamefont {H{\"a}nsch}},\ }\bibfield
  {title} {{\selectlanguage {en}\bibinfo {title} {Adaptive real-time dual-comb
  spectroscopy}},\ }\href {https://doi.org/10.1038/ncomms4375} {\bibfield
  {journal} {\bibinfo  {journal} {Nature Communications}\ }\textbf {\bibinfo
  {volume} {5}},\ \bibinfo {pages} {1} (\bibinfo {year} {2014})},\ \Eprint
  {https://arxiv.org/abs/1201.4177} {arXiv:1201.4177} \BibitemShut {NoStop}%
\bibitem [{\citenamefont {Burghoff}\ \emph {et~al.}(2016)\citenamefont
  {Burghoff}, \citenamefont {Yang},\ and\ \citenamefont
  {Hu}}]{burghoffComputationalMultiheterodyneSpectroscopy2016}%
  \BibitemOpen
  \bibfield  {author} {\bibinfo {author} {\bibfnamefont {D.}~\bibnamefont
  {Burghoff}}, \bibinfo {author} {\bibfnamefont {Y.}~\bibnamefont {Yang}},\
  and\ \bibinfo {author} {\bibfnamefont {Q.}~\bibnamefont {Hu}},\ }\bibfield
  {title} {{\selectlanguage {en}\bibinfo {title} {Computational multiheterodyne
  spectroscopy}},\ }\href {https://doi.org/10.1126/sciadv.1601227} {\bibfield
  {journal} {\bibinfo  {journal} {Science Advances}\ }\textbf {\bibinfo
  {volume} {2}},\ \bibinfo {pages} {e1601227} (\bibinfo {year}
  {2016})}\BibitemShut {NoStop}%
\bibitem [{\citenamefont {H{\'e}bert}\ \emph {et~al.}(2017)\citenamefont
  {H{\'e}bert}, \citenamefont {Genest}, \citenamefont {Desch{\^e}nes},
  \citenamefont {Bergeron}, \citenamefont {Chen}, \citenamefont {Khurmi},\ and\
  \citenamefont {Lancaster}}]{hebertSelfcorrectedChipbasedDualcomb2017}%
  \BibitemOpen
  \bibfield  {author} {\bibinfo {author} {\bibfnamefont {N.~B.}\ \bibnamefont
  {H{\'e}bert}}, \bibinfo {author} {\bibfnamefont {J.}~\bibnamefont {Genest}},
  \bibinfo {author} {\bibfnamefont {J.-D.}\ \bibnamefont {Desch{\^e}nes}},
  \bibinfo {author} {\bibfnamefont {H.}~\bibnamefont {Bergeron}}, \bibinfo
  {author} {\bibfnamefont {G.~Y.}\ \bibnamefont {Chen}}, \bibinfo {author}
  {\bibfnamefont {C.}~\bibnamefont {Khurmi}},\ and\ \bibinfo {author}
  {\bibfnamefont {D.~G.}\ \bibnamefont {Lancaster}},\ }\bibfield  {title}
  {{\selectlanguage {en}\bibinfo {title} {Self-corrected chip-based dual-comb
  spectrometer}},\ }\href {https://doi.org/10.1364/OE.25.008168} {\bibfield
  {journal} {\bibinfo  {journal} {Optics Express}\ }\textbf {\bibinfo {volume}
  {25}},\ \bibinfo {pages} {8168} (\bibinfo {year} {2017})}\BibitemShut
  {NoStop}%
\bibitem [{\citenamefont {Sterczewski}\ \emph
  {et~al.}(2019{\natexlab{a}})\citenamefont {Sterczewski}, \citenamefont
  {Westberg},\ and\ \citenamefont
  {Wysocki}}]{sterczewskiComputationalCoherentAveraging2019}%
  \BibitemOpen
  \bibfield  {author} {\bibinfo {author} {\bibfnamefont {L.~A.}\ \bibnamefont
  {Sterczewski}}, \bibinfo {author} {\bibfnamefont {J.}~\bibnamefont
  {Westberg}},\ and\ \bibinfo {author} {\bibfnamefont {G.}~\bibnamefont
  {Wysocki}},\ }\bibfield  {title} {{\selectlanguage {EN}\bibinfo {title}
  {Computational coherent averaging for free-running dual-comb spectroscopy}},\
  }\href {https://doi.org/10.1364/OE.27.023875} {\bibfield  {journal} {\bibinfo
   {journal} {Optics Express}\ }\textbf {\bibinfo {volume} {27}},\ \bibinfo
  {pages} {23875} (\bibinfo {year} {2019}{\natexlab{a}})}\BibitemShut {NoStop}%
\bibitem [{\citenamefont {Sterczewski}\ \emph
  {et~al.}(2019{\natexlab{b}})\citenamefont {Sterczewski}, \citenamefont
  {Przew{\l}oka}, \citenamefont {Kaszub},\ and\ \citenamefont
  {Sotor}}]{sterczewskiComputationalDopplerlimitedDualcomb2019}%
  \BibitemOpen
  \bibfield  {author} {\bibinfo {author} {\bibfnamefont {{\L}.~A.}\
  \bibnamefont {Sterczewski}}, \bibinfo {author} {\bibfnamefont
  {A.}~\bibnamefont {Przew{\l}oka}}, \bibinfo {author} {\bibfnamefont
  {W.}~\bibnamefont {Kaszub}},\ and\ \bibinfo {author} {\bibfnamefont
  {J.}~\bibnamefont {Sotor}},\ }\bibfield  {title} {\bibinfo {title}
  {Computational {{Doppler}}-limited dual-comb spectroscopy with a free-running
  all-fiber laser},\ }\href {https://doi.org/10.1063/1.5117847} {\bibfield
  {journal} {\bibinfo  {journal} {APL Photonics}\ }\textbf {\bibinfo {volume}
  {4}},\ \bibinfo {pages} {116102} (\bibinfo {year} {2019}{\natexlab{b}})},\
  \Eprint {https://arxiv.org/abs/1905.04647} {arXiv:1905.04647} \BibitemShut
  {NoStop}%
\bibitem [{\citenamefont {Burghoff}\ \emph {et~al.}(2019)\citenamefont
  {Burghoff}, \citenamefont {Han},\ and\ \citenamefont
  {Shin}}]{burghoffGeneralizedMethodComputational2019}%
  \BibitemOpen
  \bibfield  {author} {\bibinfo {author} {\bibfnamefont {D.}~\bibnamefont
  {Burghoff}}, \bibinfo {author} {\bibfnamefont {N.}~\bibnamefont {Han}},\ and\
  \bibinfo {author} {\bibfnamefont {J.~H.}\ \bibnamefont {Shin}},\ }\bibfield
  {title} {{\selectlanguage {EN}\bibinfo {title} {Generalized method for the
  computational phase correction of arbitrary dual comb signals}},\ }\href
  {https://doi.org/10.1364/OL.44.002966} {\bibfield  {journal} {\bibinfo
  {journal} {Optics Letters}\ }\textbf {\bibinfo {volume} {44}},\ \bibinfo
  {pages} {2966} (\bibinfo {year} {2019})}\BibitemShut {NoStop}%
\bibitem [{\citenamefont {Pinkowski}\ \emph {et~al.}(2020)\citenamefont
  {Pinkowski}, \citenamefont {Ding}, \citenamefont {Strand}, \citenamefont
  {Hanson}, \citenamefont {Horvath},\ and\ \citenamefont
  {Geiser}}]{pinkowskiDualcombSpectroscopyHightemperature2020}%
  \BibitemOpen
  \bibfield  {author} {\bibinfo {author} {\bibfnamefont {N.~H.}\ \bibnamefont
  {Pinkowski}}, \bibinfo {author} {\bibfnamefont {Y.}~\bibnamefont {Ding}},
  \bibinfo {author} {\bibfnamefont {C.~L.}\ \bibnamefont {Strand}}, \bibinfo
  {author} {\bibfnamefont {R.~K.}\ \bibnamefont {Hanson}}, \bibinfo {author}
  {\bibfnamefont {R.}~\bibnamefont {Horvath}},\ and\ \bibinfo {author}
  {\bibfnamefont {M.}~\bibnamefont {Geiser}},\ }\bibfield  {title}
  {{\selectlanguage {en}\bibinfo {title} {Dual-comb spectroscopy for
  high-temperature reaction kinetics}},\ }\href
  {https://doi.org/10.1088/1361-6501/ab6ecc} {\bibfield  {journal} {\bibinfo
  {journal} {Measurement Science and Technology}\ }\textbf {\bibinfo {volume}
  {31}},\ \bibinfo {pages} {055501} (\bibinfo {year} {2020})},\ \Eprint
  {https://arxiv.org/abs/1903.07578} {arXiv:1903.07578} \BibitemShut {NoStop}%
\bibitem [{\citenamefont {Karpov}\ \emph {et~al.}(2018)\citenamefont {Karpov},
  \citenamefont {Pfeiffer}, \citenamefont {Liu}, \citenamefont {Lukashchuk},\
  and\ \citenamefont {Kippenberg}}]{karpovPhotonicChipbasedSoliton2018}%
  \BibitemOpen
  \bibfield  {author} {\bibinfo {author} {\bibfnamefont {M.}~\bibnamefont
  {Karpov}}, \bibinfo {author} {\bibfnamefont {M.~H.~P.}\ \bibnamefont
  {Pfeiffer}}, \bibinfo {author} {\bibfnamefont {J.}~\bibnamefont {Liu}},
  \bibinfo {author} {\bibfnamefont {A.}~\bibnamefont {Lukashchuk}},\ and\
  \bibinfo {author} {\bibfnamefont {T.~J.}\ \bibnamefont {Kippenberg}},\
  }\bibfield  {title} {{\selectlanguage {en}\bibinfo {title} {Photonic
  chip-based soliton frequency combs covering the biological imaging window}},\
  }\href {https://doi.org/10.1038/s41467-018-03471-x} {\bibfield  {journal}
  {\bibinfo  {journal} {Nature Communications}\ }\textbf {\bibinfo {volume}
  {9}},\ \bibinfo {pages} {1} (\bibinfo {year} {2018})}\BibitemShut {NoStop}%
\bibitem [{\citenamefont {Klocke}\ \emph {et~al.}(2018)\citenamefont {Klocke},
  \citenamefont {Mangold}, \citenamefont {Allmendinger}, \citenamefont {Hugi},
  \citenamefont {Geiser}, \citenamefont {Jouy}, \citenamefont {Faist},\ and\
  \citenamefont {Kottke}}]{klockeSingleShotSubmicrosecondMidinfrared2018}%
  \BibitemOpen
  \bibfield  {author} {\bibinfo {author} {\bibfnamefont {J.~L.}\ \bibnamefont
  {Klocke}}, \bibinfo {author} {\bibfnamefont {M.}~\bibnamefont {Mangold}},
  \bibinfo {author} {\bibfnamefont {P.}~\bibnamefont {Allmendinger}}, \bibinfo
  {author} {\bibfnamefont {A.}~\bibnamefont {Hugi}}, \bibinfo {author}
  {\bibfnamefont {M.}~\bibnamefont {Geiser}}, \bibinfo {author} {\bibfnamefont
  {P.}~\bibnamefont {Jouy}}, \bibinfo {author} {\bibfnamefont {J.}~\bibnamefont
  {Faist}},\ and\ \bibinfo {author} {\bibfnamefont {T.}~\bibnamefont
  {Kottke}},\ }\bibfield  {title} {\bibinfo {title} {Single-{{Shot
  Sub}}-microsecond {{Mid}}-infrared {{Spectroscopy}} on {{Protein Reactions}}
  with {{Quantum Cascade Laser Frequency Combs}}},\ }\href
  {https://doi.org/10.1021/acs.analchem.8b02531} {\bibfield  {journal}
  {\bibinfo  {journal} {Analytical Chemistry}\ }\textbf {\bibinfo {volume}
  {90}},\ \bibinfo {pages} {10494} (\bibinfo {year} {2018})}\BibitemShut
  {NoStop}%
\bibitem [{\citenamefont {Hoffmann}\ and\ \citenamefont
  {Russer}(2011)}]{hoffmannRealTimeLowNoiseUltrabroadband2011}%
  \BibitemOpen
  \bibfield  {author} {\bibinfo {author} {\bibfnamefont {C.}~\bibnamefont
  {Hoffmann}}\ and\ \bibinfo {author} {\bibfnamefont {P.}~\bibnamefont
  {Russer}},\ }\bibfield  {title} {\bibinfo {title} {A {{Real}}-{{Time
  Low}}-{{Noise Ultrabroadband Time}}-{{Domain EMI Measurement System}} up to
  18 {{GHz}}},\ }\href {https://doi.org/10.1109/TEMC.2011.2158827} {\bibfield
  {journal} {\bibinfo  {journal} {IEEE Transactions on Electromagnetic
  Compatibility}\ }\textbf {\bibinfo {volume} {53}},\ \bibinfo {pages} {882}
  (\bibinfo {year} {2011})}\BibitemShut {NoStop}%
\end{thebibliography}%

\end{document}